\documentclass[printer]{aa}
\usepackage{graphicx}
\def\lesssim{\mathrel{\hbox{\rlap{\hbox{\lower4pt\hbox{$\sim$}}}\hbox{$<$}}}}
\def\gtrsim{\mathrel{\hbox{\rlap{\hbox{\lower4pt\hbox{$\sim$}}}\hbox{$>$}}}}
\newcommand{\mincir}{\raise -2.truept\hbox{\rlap{\hbox{$\sim$}}\raise5.truept
\hbox{$<$}\ }}
\newcommand{\magcir}{\raise -2.truept\hbox{\rlap{\hbox{$\sim$}}\raise5.truept
\hbox{$>$}\ }}
\newcommand{\siml}{\raise -2.truept\hbox{\rlap{\hbox{$\sim$}}\raise5.truept
\hbox{$<$}\ }}
\newcommand{\simg}{\raise -2.truept\hbox{\rlap{\hbox{$\sim$}}\raise5.truept
\hbox{$>$}\ }}
\newcommand{\be}{\begin{equation}}
\newcommand{\ee}{\end{equation}}
\newcommand{\ba}{\begin{eqnarray}}
\newcommand{\ea}{\end{eqnarray}}
\newcommand {\h} {$h^{-1}$ Mpc $ \;$}
\newcommand {\kpc} {$h^{-1}$ kpc}
\newcommand {\hh} {$h^{-1}$ Mpc}
\newcommand {\ks} {km~s$^{-1} \;$}
\newcommand {\kss} {km~s$^{-1}$}
\newcommand {\msun} {$h^{-1} \ M_{\odot} \;$}

\begin{document}
   \title{Internal dynamics of the radio--halo cluster A2219: \\ a
multi--wavelength analysis \thanks{Based on observations made on the
island of La Palma with the Italian Telescopio Nazionale Galileo (TNG)
operated by the Centro Galileo Galilei of the INAF (Istituto Nazionale
di Astrofisica) and with the 1.0m Jacobus Kapteyn Telescope (JKT)
operated by the Isaac Newton Group at the Spanish Observatorio de
Roque de los Muchachos of the Instituto de Astrofisica de Canarias.}}
%
\author{W. Boschin\inst{1} 
\and M. Girardi\inst{1}
\and R. Barrena\inst{2,3}
\and A. Biviano\inst{4}
\and L. Feretti\inst{5}
\and M. Ramella\inst{4}}
   \offprints{W. Boschin}

\institute{Dipartimento di Astronomia, Universit\`{a} degli Studi di
Trieste, via Tiepolo 11, 34131 Trieste, Italy\\
\email{boschin,girardi@ts.astro.it} \and Instituto de Astrofisica
de Canarias, E-38200 La Laguna, Tenerife, Spain \\ \email{rbarrena@ll.iac.es} \and INAF - Telescopio
Nazionale Galileo, Roque de Los Muchachos, PO box 565, 38700 Santa
Cruz de La Palma, Spain \and INAF -
Osservatorio Astronomico di Trieste, via Tiepolo 11, 34131 Trieste,
Italy\\ \email{biviano,ramella@ts.astro.it} \and Istituto di Radioastronomia
del C.N.R., via Gobetti 101, 40129 Bologna, Italy\\
\email{lferetti@ira.cnr.it} } \date{Received / Accepted}

\abstract{We present the results of the dynamical analysis of the
rich, hot, and X--ray very luminous galaxy cluster A2219, containing a
powerful diffuse radio--halo.  Our analysis is based on new redshift
data for 27 galaxies in the cluster region, measured from spectra
obtained at the TNG, with the addition of other 105 galaxies recovered
from reduction of CFHT archive data in a cluster region of $\sim
5\arcmin$ radius ($\sim 0.8$ \h at the cluster distance) centered on
the cD galaxy.  The investigation of the dynamical status is also
performed by using X--ray data stored in the Chandra archive.
Further, valuable information comes from other bands -- optical
photometric, infrared, and radio data -- which are analyzed and/or
discussed, too.  We find that A2219 appears as a peak in the velocity
space at $z=0.225$, and select 113 cluster members. We compute a high
value for the line--of--sight velocity dispersion, $\sigma_{\rm v}=
1438^{+109}_{-86}$ \kss, consistent with the high average X--ray
temperature of 10.3 keV. If dynamical equilibrium is assumed, the
virial theorem leads to $M\sim2.8\times 10^{15}$\msun for the global
mass within the virial region.  However, further investigation based
on both optical and X--ray data shows significant signs of a young
dynamical status. In fact, we find strong evidence for the elongation
of the cluster in the SE--NW direction coupled with a significant
velocity gradient, as well as for the presence of substructure both in
optical data and X--ray data.  Moreover, we point out the presence of
several active galaxies. We discuss the results of our
multi--wavelength investigation suggesting a complex merging scenario
where the main, original structure is subject to an ongoing merger
with few clumps aligned in a filament in the foreground oriented along
an oblique direction with respect to the line--of--sight. Our
conclusion supports the view of the connection between extended radio
emission and merging phenomena in galaxy clusters.

}
\authorrunning{Boschin et al.}
\titlerunning{Internal dynamics of A2219} 
\maketitle
%

\section{Introduction}

In the hierarchical scenario for large--scale--structure formation,
mergers are an essential ingredient of galaxy cluster evolution (e.g.,
White \cite{whi97}; Evrard \& Gioia \cite{evr02}).

One of the most recent aspects of the merging phenomenology is the
possible connection of cluster mergers with the presence of extended,
diffuse radio sources, halos and relics (Feretti \cite{fer99}; Buote
\cite{buo02}; Giovannini \& Feretti \cite{gio02}).  Radio--halos are
located at the cluster center, with typical sizes of 0.5 $h^{-1}$ Mpc,
regular shape, steep radio spectra and no significant polarization;
while relics are located in peripheral regions of the clusters, with
irregular shape and generally highly polarized spectra (e.g., Feretti
\& Giovannini \cite{fer96}; Giovannini \& Feretti \cite{gio02}). The
synchrotron radio emission of halos and relics demonstrates the
existence of large scale cluster magnetic fields, of the order of
0.1--1 $\mu$G, and of widespread relativistic particles of energy
density 10$^{-14}$ -- 10$^{-13}$ erg cm$^{-3}$. Although the
properties of halos and relics are not well understood, several
suggestions for the mechanism transferring energy to the relativistic
electrons have been made (e.g., Ensslin \cite{ens00}).

Cluster mergers were suggested to provide the large amount of energy
necessary for electron re--acceleration and magnetic field amplification
(Feretti \cite{fer99}; Feretti \cite{fer02}; Sarazin
\cite{sar02}). However, the precise radio--halo formation scenario is
still debated (e.g., Ensslin \& R\"ottgering \cite{ens02}).  In fact,
the diffuse radio sources are quite uncommon and only recently we can
study these phenomena on the basis of a sufficient statistics ($\sim
30$ clusters up to $z\sim 0.3$, e.g., Giovannini et al. \cite{gio99};
see also Giovannini \& Feretti \cite{gio02}).

Presently, growing evidence of connection between diffuse emission and
cluster merging is based on X--ray data (e.g., B\"ohringer \&
Schuecker \cite{boh02}; Buote \cite{buo02}). Studies based
on a large number of clusters have found a significant relation
between the radio and the X--ray surface brightness (Govoni et
al. \cite{gov01}) and connections between the presence of
radio--halos/relics and irregular and bimodal X--ray surface
brightness distribution (Schuecker et al. \cite{sch01}).  However, we
are far from understanding the whole picture. Unfortunately, to date
optical data is lacking or poorly exploited. Sparse literature
concerns some few individual clusters (e.g., Colless \& Dunn \cite{col96};
G\'omez et al. \cite{gom00}; Barrena et al. \cite{bar02}; Mercurio et
al. \cite{mer03}).

Optical data are a powerful way to investigate the presence and the
dynamics of cluster mergers (e.g., Girardi \& Biviano \cite{gir02}),
too.  The spatial and kinematic analysis of member galaxies allow us
to detect and measure the amount of substructure, to identify and
analyze possible pre--merging clumps or merger remnants.  This optical
information is really complementary to X--ray information since
galaxies and ICM react on different time scales during a merger (see
numerical simulations by Roettiger et al. \cite{roe97}). Moreover,
additional information comes from spectral types of member galaxies,
since cluster mergers could stimulate star formation in cluster
galaxies; thus the spectral signatures of past activity are useful to
determine the relevant time--scales (e.g., Bekki \cite{bek99};
Terlevich et al. \cite{ter99}).

To investigate the connection between diffuse radio sources and
cluster mergers, we have performed spectroscopic observations at TNG
and a multi--wavelength analysis of the Abell cluster A2219 (Abell et
al. \cite{abe89}). This cluster shows the presence of a powerful
radio--halo (Giovannini et al. \cite{gio99}; Bacchi et
al. \cite{bac03}) and is a rich, X--ray luminous, hot cluster at
moderate redshift $z\sim0.22$ -- Abell richness $=3$ (Abell et
al. \cite{abe89}); $L_\mathrm{X}$(0.1--2.4 keV)$\sim$5$\times
10^{44} \ h^{-2}$ erg\ s$^{-1}$ (Ebeling et al. \cite{ebe96});
$T_\mathrm{X}\sim$10 keV (Rizza et al. \cite{riz98}). Possible
evidence of the young dynamical status of A2219 comes from the
ROSAT/HRI X--ray data analysis and strong/weak lensing features (Rizza
et al. \cite{riz98}; Smail et al. \cite{sma95}; Dahle et
al. \cite{dah02}).

This paper is organized as follows.  We present the new redshift data
of A2219 in Sect.~2 and the analysis of optical data in Sect.~3. Our
analysis of recent X--ray Chandra archival data is shown in Sect.~4.
Active galaxies are analyzed and discussed on the basis of
multi--wavelength data using also mid--IR and radio data in
Sect.~5. Finally, in Sect.~6, we summarize and discuss our results
suggesting a tentative picture of the dynamical status of A2219.

Unless otherwise stated, we give errors at the 68\% confidence
level (hereafter c.l.)

Throughout the paper, we assume a flat cosmology with $\Omega_{\rm m}=0.3$
and $\Omega_{\Lambda}=0.7$.  For the sake of simplicity in rescaling
we adopt a Hubble constant of 100 \ks \hh.  In this context, 1$\arcmin$
corresponds to $\sim 0.152$ \hh.

\section{Optical observations and data reduction}

Multi--object spectroscopic observations of A2219 were carried out at
the TNG telescope in May 2003 during the program of proposal AOT--7
ID:18. We used DOLORES/MOS with the LR--B Grism 1, yielding a
dispersion of 187 \AA/mm, and the Loral CCD of $2048\times2048$ pixels
(pixel size of 15 $\mu$m).  This combination of grating and detector
results in dispersions of 2.8 \AA/pix.  We have taken 1 MOS mask with
39 slits.  We acquired two exposures of 1800.0 s each.  Wavelength
calibration was done using arc lamps before or after each exposure
(Helium--Argon).

Reduction of spectroscopic data was carried out with IRAF
\footnote{IRAF is distributed by the National Optical Astronomy
Observatories, which are operated by the Association of Universities
for Research in Astronomy, Inc., under cooperative agreement with the
National Science Foundation.}. The signal-to-noise ratio per pixel was
computed from the ratio between the signal at 6000 \AA$\,\,$ and the
$rms$ noise within 4000-7000 \AA. The resulting average
signal-to-noise ratio of our spectra is S/N$\sim$16.

Radial velocities are determined using the cross--correlation
technique (Tonry \& Davis \cite{ton79}) implemented in the RVSAO
package (developed at the Smithsonian Astrophysical Observatory
Telescope Data Center).  Each spectrum is correlated against six
templates for a variety of galaxy spectral types: E, S0, Sa, Sb, Sc,
Ir (Kennicutt \cite{ken92}).  The template producing the highest value
of $\cal R$, i.e., the parameter given by RVSAO and related to the
signal--to--noise of the correlation peak, is chosen.  Moreover, all
spectra and their best correlation functions are examined visually to
verify the redshift determination.  In some ambiguous cases, generally
very late type galaxies, we use EMSAO package to confirm the estimated
redshift, and in two cases we take the EMSAO redshift as a more
reliable estimate of the redshift.

Out of the 39 spectra of objects, 9 turned out to be objects at
$z\sim 0$ (included alignment stars for masks), while for 3 cases,
with $\cal R$~$\lesssim 4$, we cannot determine the redshift. Thus, we
obtain spectra for 27 galaxies.

We added to our data observations stored in CFHT archive (proposal ID:
01AF37, P.I.: J.P. Kneib): 1 MOS mask of one 1800 s --exposure with 155
slits. We reduced the spectra with the same procedure adopted for TNG
ones.  We succeeded in obtaining spectra for 111 galaxies, 6 of which
are in common with TNG galaxies.  The average S/N of these CFHT
spectra is 8.

In order to combine our TNG data with those from CFHT archive, we have
to check for a possible zero--point correction. To this aim we use the
six galaxies with double redshift determination.  This allow us to
obtain a more rigorous estimate for the redshift errors since the
nominal errors as given by the cross--correlation are known to be
smaller than the true errors (e.g., Malamuth et al.  \cite{mal92};
Bardelli et al. \cite{bar94}; Ellingson \& Yee \cite{ell94}; Quintana
et al. \cite{qui00}). Thus, for the six galaxies with double redshift
determination, we fit the first determination vs. the second one by
using a straight line and considering errors in both coordinates
(e.g., Press et al. \cite{pre92}). The fitted line agrees with the one
to one relation (slope=$0.996\pm0.004$, intercept=$(1\pm1)\times
10^{-3}$), but the small value of $\chi^2$--probability indicates a
poor fit, suggesting that the nominal cross--correlation errors are
underestimated.  Only when nominal errors are multiplied by a $\sim
2.3$ factor the observed scatter can be explained. Therefore,
hereafter we assume that true errors are larger than nominal
cross--correlation errors by a factor 2.3. Slightly smaller correction
factors are obtained for nearby clusters (e.g., Malamuth et
al. \cite{mal92}; Bardelli et al. \cite{bar94}; Quintana et
al. \cite{qui00}).

Our spectroscopic catalogue consists of 132 galaxies sampling a cluster
region of 5$\arcmin$ around the dominant galaxy (no. 65), which is a
cD galaxy (e.g., Allen et al. \cite{all92}).

We also determined the equivalent widths (EW hereafter) of the
emission line [OII] and the absorption line H$\delta$, which are good
indicators of current and recent star formation, respectively.
Hereafter all the EW are in angstroms and are positive if the line is
in absorption and negative if the line is in emission.  We estimate
minimum measurable EW as the width of a line spanning 2.8 \AA (our
dispersion) in wavelength, with an intensity three times the rms noise
in the adjacent continuum.

A variety of spectroscopic--classification schemes are presented in
the literature (see, e.g., Couch et al. \cite{cou94}; Dressler et
al. \cite{dre99} and refs. therein). We use a conservative approach
leading to a sparse classification ($45\%$ of the sample, see
Table~\ref{tab1}).  We define $e$--type galaxies those showing strong
active star formation as indicated by the presence of an [OII] line
with an equivalent width of EW([OII])$\lesssim-15$ \AA$\,\,$(e.g.,
Hammer et al. \cite{ham97}; Postman et al. \cite{pos98}).  Out of
galaxies having S/N$> 10$ we define $a$--type, those with strong
Balmer absorption EW(H$\delta$)$> 4$ \AA$\,\,$ (e.g., the
``post--starburst'' galaxies by Couch et al. \cite{cou94}).  We find
fourteen ``active'' galaxies: four of $e$--type --- measured EW([OII])
in the range $[-39.3,-14.3]$ \AA; ten of $a$--type --- measured
EW(H$\delta$) in the range [4.0;12.4] \AA.  Out of non active galaxies
with S/N$> 10$ those (37) with measured EW(H$\delta$)$<$3 \AA$\,\,$
are classified as passive galaxies ($k$--type) and the remaining nine
galaxies, intermediate between $a$ and $k$ are classified as
$i$--type. Non active galaxies do not show significant [OII] emission
and only two $a$--galaxies show a very small [OII] emission (-3.5\AA).

As far as photometry is concerned, we obtained exposures on a
$10\arcmin\times10\arcmin$ unvignetted field centered on $\alpha =
16^{\mathrm{h}}\,40^{\mathrm{m}}\,22^{\mathrm{s}}$, $\delta =
46\degr\,42\arcmin\,15\arcsec$ (J2000) through B and R Harris filters
using the CCD camera mounted on the 1m JKT telescope (at Roque de los
Muchachos Observatory, La Palma) in July 2003. We performed several
exposures for total integration times of 3000 and 7200 s in R and B
band, respectively, with seeing conditions about 1.4$\arcsec$. We
carried out the image reduction (bias, flat and fringing corrections)
using IRAF package and photometry using AUTOMAG SExtractor procedure
(Bertin \& Arnouts \cite{ber96}). This photometry was calibrated using
Landolt's standard fields at the JKT and TNG telescopes and
transformed into the Johnson filter system. We estimate that our
photometric sample is complete down to R= 19.7 and B=21.4 within the
observed field.

We assign R magnitudes to 131 out of the 132 galaxies of our
spectroscopic catalogue: the only exception is galaxy no. 52, which is
too close to a very bright star.  Out of these 131 galaxies, 129 have
assigned magnitudes in B band, too.  We measure redshift for galaxies
down to $R \sim 20.5$ mag, but a high level of completeness is
obtained only for bright galaxies ($\sim$66\% completeness for
R$<$19.7 mag).

%

\begin{table}
        \caption[]{Velocity catalog of 132 spectroscopically measured galaxies. in Column~1, IDs in italics indicate non--cluster galaxies.}
         \label{catalogue}
              $$ 
           \begin{array}{r c c c c r c c}
            \hline
            \noalign{\smallskip}
            \hline
            \noalign{\smallskip}

\mathrm{ID} & \mathrm{\alpha},\mathrm{\delta}\,(\mathrm{J}2000)  & \mathrm{R} & \mathrm{B-R}  & \mathrm{v} & \mathrm{\Delta}\mathrm{v} & \mathrm{SC} & \mathrm{Tel.}\\
  & 16^h      , +46^o    &  &  &\,\,\,\,\,\,\,\mathrm{(\,km}&\mathrm{s^{-1}\,)}\,\,\,&  &\\
            \hline
            \noalign{\smallskip}   

  1             &     39\ 54.53 , 40\ 16.4 & 19.89 &   2.15 &  67614 & 223 &   & \mathrm{C}\\
  2             &     39\ 55.54 , 41\ 01.4 & 18.80 &   1.80 &  68948 & 117 & a & \mathrm{C}\\
  \textit{3}    &     39\ 55.94 , 42\ 28.7 & 17.89 &   1.11 &  33327 & 205 & k & \mathrm{C}\\
  4             &     39\ 56.69 , 41\ 40.4 & 18.89 &   2.26 &  67403 &  97 &   & \mathrm{C}\\
  5             &     39\ 56.76 , 43\ 35.4 & 18.19 &   2.34 &  68244 &  87 & k & \mathrm{C}\\
  \textit{6}    &     39\ 56.88 , 40\ 24.1 & 18.33 &   2.11 &  53453 &  71 & k & \mathrm{C}\\
  7             &     39\ 59.26 , 41\ 42.0 & 19.83 &   2.11 &  69556 & 248 &   & \mathrm{C}\\
  8             &     39\ 59.66 , 44\ 24.0 & 18.80 &   1.95 &  67681 &  85 & i & \mathrm{C}\\
  9             &     40\ 01.63 , 41\ 50.6 & 19.68 &   1.62 &  69022 & 198 &   & \mathrm{C}\\
 10             &     40\ 02.78 , 44\ 15.3 & 19.73 &   2.14 &  67205 & 198 &   & \mathrm{C}\\
 11             &     40\ 03.19 , 41\ 53.7 & 18.79 &   2.19 &  68023 & 152 &   & \mathrm{C}\\
 12             &     40\ 03.36 , 43\ 58.8 & 18.74 &   2.43 &  66980 & 108 &   & \mathrm{C}\\
 13             &     40\ 03.50 , 41\ 38.0 & 19.91 &   1.70 &  68235 &  92 & a & \mathrm{T}\\
 14             &     40\ 03.70 , 46\ 37.5 & 18.07 &   2.20 &  70167 & 136 & i & \mathrm{C}\\
 15             &     40\ 03.91 , 40\ 45.8 & 20.07 &   2.35 &  67995 &  85 & k & \mathrm{T}\\
 16             &     40\ 04.06 , 44\ 05.7 & 19.57 &   2.20 &  67483 & 145 &   & \mathrm{C}\\
 17             &     40\ 04.10 , 46\ 45.2 & 18.27 &   1.81 &  69424 & 244 & i & \mathrm{C}\\
 \textit{18}    &     40\ 04.15 , 40\ 57.4 & 19.43 &   1.09 &  33073 & 191 & k & \mathrm{T}\\
 19             &     40\ 04.39 , 42\ 08.6 & 18.46 &   2.18 &  64785 &  94 &   & \mathrm{C}\\
 20             &     40\ 04.92 , 42\ 00.2 & 19.80 &   1.65 &  67906 & 136 &   & \mathrm{C}\\
 21             &     40\ 05.93 , 42\ 04.0 & 19.14 &   1.39 &  67488 & 156 &   & \mathrm{C}\\
 22             &     40\ 06.19 , 44\ 19.5 & 19.40 &   2.06 &  67739 & 106 &   & \mathrm{C}\\
 \textit{23}    &     40\ 06.46 , 46\ 25.9 & 17.53 &   1.85 &  43521 & 138 & k & \mathrm{C}\\
 24             &     40\ 06.72 , 40\ 40.2 & 18.21 &   2.31 &  68595 &  69 &   & \mathrm{C}\\
 25             &     40\ 07.15 , 41\ 06.5 & 18.53 &   1.99 &  65555 & 156 &   & \mathrm{C}\\
 26             &     40\ 07.27 , 45\ 30.5 & 18.92 &   2.11 &  68483 & 106 &   & \mathrm{C}\\
 \textit{27}    &     40\ 08.66 , 41\ 42.9 & 19.52 &   1.52 &  57266 & 163 & i & \mathrm{C}\\
 28             &     40\ 08.81 , 43\ 40.6 & 18.78 &   2.37 &  69612 &  60 & k & \mathrm{C}\\
 29             &     40\ 09.12 , 45\ 48.7 & 19.33 &   1.60 &  68317 &  92 &   & \mathrm{C}\\
 30             &     40\ 09.14 , 41\ 52.5 & 19.54 &   2.07 &  65020 & 133 &   & \mathrm{C}\\
 31             &     40\ 09.17 , 42\ 11.3 & 19.98 &   2.02 &  70704 & 108 &   & \mathrm{T}\\
 32             &     40\ 09.46 , 43\ 32.8 & 18.99 &   1.58 &  67772 & 122 & a & \mathrm{C}\\
 33             &     40\ 09.65 , 43\ 03.2 & 19.86 &   2.33 &  67773 &  81 & k & \mathrm{T}\\
 34             &     40\ 10.20 , 45\ 32.1 & 18.91 &   2.35 &  70238 & 143 &   & \mathrm{C}\\
 35             &     40\ 10.63 , 43\ 22.5 & 18.74 &   2.08 &  70212 &  83 & a & \mathrm{C}\\
 36             &     40\ 11.47 , 42\ 31.2 & 19.74 &   2.18 &  70824 &  90 & k & \mathrm{T}\\
 37             &     40\ 11.52 , 43\ 19.8 & 18.61 &   2.30 &  67655 &  83 & k & \mathrm{C}\\
 \textit{38}    &     40\ 11.76 , 38\ 53.9 & 20.06 &   1.61 & 106730 & 168 & e & \mathrm{T}\\
 39             &     40\ 12.53 , 41\ 59.6 & 19.41 &   2.09 &  66638 &  62 & k & \mathrm{T}\\
 40             &     40\ 12.55 , 45\ 35.0 & 20.12 &   1.97 &  66657 & 129 &   & \mathrm{C}\\
 41             &     40\ 12.74 , 41\ 10.6 & 18.02 &   2.11 &  70175 &  64 & k & \mathrm{C}\\
              \noalign{\smallskip}			    
            \hline					    
            \noalign{\smallskip}			    
            \hline					    
         \end{array}
     $$ 
         \end{table}
\addtocounter{table}{-1}
\begin{table}
          \caption[ ]{Continued.}
     $$ 
           \begin{array}{r c c c c r c c}
            \hline
            \noalign{\smallskip}
            \hline
            \noalign{\smallskip}

\mathrm{ID} & \mathrm{\alpha},\mathrm{\delta}\,(\mathrm{J}2000)  & \mathrm{R} & \mathrm{B-R}  & \mathrm{v} & \mathrm{\Delta}\mathrm{v} & \mathrm{SC} & \mathrm{Tel.}\\
  & 16^h      , +46^o    &  &  &\,\,\,\,\,\,\,\mathrm{(\,km}&\mathrm{s^{-1}\,)}\,\,\,&  &\\
            \hline
            \noalign{\smallskip}
 42             &     40\ 13.10 , 43\ 06.2 & 18.82 &   1.26 &  69084 &  69 & a,I & \mathrm{C}\\
 43             &     40\ 13.18 , 44\ 40.5 & 18.13 &   2.21 &  67154 &  67 & i & \mathrm{C}\\
 44             &     40\ 13.80 , 44\ 50.6 & 19.81 &  2.13 & 67068 & 140 &   & \mathrm{C}\\
 \textit{45}    &     40\ 14.04 , 41\ 49.5 & 17.44 &  1.41 & 33192 &  97 & k,I & \mathrm{C}\\
 46             &     40\ 14.26 , 43\ 04.9 & 18.89 &  2.22 & 67587 &  83 &   & \mathrm{C}\\
 47             &     40\ 14.35 , 43\ 52.9 & 19.44 &  1.73 & 66045 &  94 & k & \mathrm{T}\\
 48             &     40\ 14.42 , 46\ 14.7 & 18.84 &  2.13 & 65166 & 117 &   & \mathrm{C}\\
 49             &     40\ 14.64 , 43\ 04.4 & 19.96 &  1.98 & 67437 & 163 &   & \mathrm{C}\\
 50             &     40\ 15.70 , 45\ 48.6 & 19.38 &  2.32 & 69168 & 143 &   & \mathrm{C}\\
 51             &     40\ 15.89 , 42\ 31.5 & 18.54 &  2.11 & 68411 &  85 & i,R & \mathrm{C}\\
 52             &     40\ 16.10 , 44\ 12.4 & ..... &  .... & 69563 &  78 & i & \mathrm{C}\\
 53             &     40\ 16.20 , 40\ 20.8 & 20.21 &  1.65 & 68334 & 110 & a & \mathrm{T}\\
 54             &     40\ 16.66 , 46\ 06.7 & 19.06 &  2.30 & 69540 & 186 &   & \mathrm{C}\\
 55             &     40\ 16.68 , 43\ 09.4 & 20.35 &  2.20 & 67937 & 255 &   & \mathrm{C}\\
 56             &     40\ 16.92 , 42\ 47.9 & 19.98 &  2.33 & 68319 &  90 & k & \mathrm{T}\\ 
 57             &     40\ 16.97 , 39\ 50.7 & 20.38 &  0.39 & 64664 & 186 &   & \mathrm{C}\\
 58             &     40\ 17.18 , 42\ 11.3 & 19.03 &  1.88 & 66799 & 147 &   & \mathrm{C}\\
 59             &     40\ 17.95 , 44\ 38.4 & 19.81 &  1.85 & 67744 & 108 &   & \mathrm{C}\\
 60             &     40\ 18.19 , 42\ 36.0 & 18.28 &  2.70 & 69246 &  97 &   & \mathrm{C}\\
 \textit{61}    &     40\ 18.36 , 40\ 21.3 & 17.30 &  1.71 & 56619 &  62 & i & \mathrm{C}\\
 62             &     40\ 19.13 , 42\ 41.5 & 19.24 &  1.26 & 68558 & 168 &   & \mathrm{C}\\
 63             &     40\ 19.25 , 40\ 08.8 & 19.10 &  1.24 & 68590 &  51 & k & \mathrm{T}\\
 64             &     40\ 19.39 , 41\ 05.5 & 19.07 &  2.16 & 66975 &  90 &   & \mathrm{C}\\
 65             &     40\ 19.87 , 42\ 41.3 & 16.19 &  2.10 & 67259 &  71 & k & \mathrm{T}\\
 66             &     40\ 19.97 , 38\ 41.3 & 18.48 &  2.27 & 68681 & 106 &   & \mathrm{C}\\
 \textit{67}    &     40\ 20.38 , 40\ 12.5 & 18.13 &  1.72 & 57334 &  64 & k & \mathrm{C}\\
 68             &     40\ 20.57 , 41\ 48.7 & 18.47 &  2.35 & 70232 &  51 & k & \mathrm{T}\\
 69             &     40\ 20.70 , 45\ 51.7 & 19.47 &  2.29 & 67345 & 140 &   & \mathrm{C}\\
 70             &     40\ 20.81 , 42\ 43.9 & 18.69 &  1.54 & 66330 &  74 &   & \mathrm{C}\\
 71             &     40\ 21.50 , 42\ 39.3 & 17.78 &  1.98 & 65490 &  99 &   & \mathrm{C}\\
 \textit{72}    &     40\ 21.53 , 40\ 48.4 & 18.41 &  1.46 & 56477 & 161 & I  & \mathrm{C}\\
 \textit{73}    &     40\ 21.86 , 41\ 16.5 & 18.36 &  1.80 & 55832 &  90 & k & \mathrm{T}\\
 74             &     40\ 21.86 , 43\ 58.4 & 18.87 &  2.15 & 66947 &  74 & k & \mathrm{T}\\
 75             &     40\ 21.89 , 44\ 44.3 & 18.44 &  2.09 & 64697 &  60 & k & \mathrm{C}\\
 76             &     40\ 21.96 , 46\ 17.9 & 19.06 &  1.72 & 64878 & 108 & a & \mathrm{C}\\
 77             &     40\ 22.08 , 42\ 46.0 & 17.42 &  2.23 & 62364 &  99 & k,R & \mathrm{C}\\
 78             &     40\ 22.08 , 43\ 07.6 & 19.67 &  2.15 & 67529 &  97 & k & \mathrm{T}\\
 79             &     40\ 22.20 , 45\ 56.4 & 19.82 &  2.03 & 65624 & 122 & k & \mathrm{T}\\
 80             &     40\ 22.80 , 39\ 52.0 & 20.61 &  2.88 & 64953 & 173 &   & \mathrm{C}\\
 81             &     40\ 22.90 , 42\ 17.9 & 17.83 &  1.64 & 65516 &  83 & a,I & \mathrm{C}\\
 \textit{82}    &     40\ 23.30 , 46\ 33.0 & 16.20 &  1.45 & 32860 &  97 & k & \mathrm{C}\\
 83             &     40\ 23.40 , 44\ 53.0 & 20.71 &  1.72 & 68327 & 179 &   & \mathrm{T}\\
              \noalign{\smallskip}			    
            \hline					    
            \noalign{\smallskip}			    
            \hline					    
         \end{array}
     $$ 
         \end{table}
\addtocounter{table}{-1}
\begin{table}
          \caption[ ]{Continued.}
     $$ 
           \begin{array}{r c c c c r c c}
            \hline
            \noalign{\smallskip}
            \hline
            \noalign{\smallskip}

\mathrm{ID} & \mathrm{\alpha},\mathrm{\delta}\,(\mathrm{J}2000)  & \mathrm{R} & \mathrm{B-R}  & \mathrm{v} & \mathrm{\Delta}\mathrm{v} & \mathrm{SC} & \mathrm{Tel.}\\
  & 16^h      , +46^o    &  &  &\,\,\,\,\,\,\,\mathrm{(\,km}&\mathrm{s^{-1}\,)}\,\,\,&  &\\
            \hline
            \noalign{\smallskip}
 84             &     40\ 23.60 , 40\ 32.0 & 18.93 &  2.29 & 69604 &  74 &   & \mathrm{C}\\
 85             &     40\ 23.70 , 42\ 10.0 & 17.16 &  2.45 & 68579 &  85 & k,R & \mathrm{C}\\
 86             &     40\ 23.70 , 43\ 05.0 & 18.23 &  2.38 & 69673 &  83 & k & \mathrm{T}\\
 87             &     40\ 23.80 , 43\ 20.0 & 18.88 &   2.37 &  66376  &  60 & k & \mathrm{T}\\
 88             &     40\ 24.55 , 42\ 22.1 & 18.94 &   2.22 &  66053  &  78 & k & \mathrm{C}\\
 89             &     40\ 24.79 , 40\ 38.0 & 20.79 &   1.81 &  69885  & 150 &   & \mathrm{C}\\
 \textit{90}    &     40\ 25.01 , 44\ 18.9 & 18.28 &   1.52 &  40362  & 136 &   & \mathrm{C}\\
 91             &     40\ 25.39 , 44\ 10.5 & 19.60 &   2.30 &  67633  & 154 &   & \mathrm{C}\\
 92             &     40\ 25.54 , 43\ 42.8 & 18.71 &   2.13 &  66346  &  69 & k & \mathrm{T}\\
 93             &     40\ 25.80 , 42\ 34.7 & 17.99 &   2.10 &  63015  &  81 & i & \mathrm{C}\\
 94             &     40\ 25.87 , 40\ 34.1 & 18.66 &   1.52 &  68027  & 145 &   & \mathrm{C}\\
 95             &     40\ 26.64 , 44\ 13.9 & 19.60 &   2.24 &  67488  & 110 &   & \mathrm{C}\\
 96             &     40\ 26.95 , 45\ 43.7 & 20.64 &   1.78 &  70934  & 235 &   & \mathrm{C}\\
 97             &     40\ 27.65 , 46\ 23.1 & 18.85 &   2.23 &  67814  &  92 &   & \mathrm{C}\\
 98             &     40\ 28.30 , 41\ 46.4 & 19.81 &   2.27 &  66766  & 106 & k & \mathrm{T}\\
 99             &     40\ 28.51 , 46\ 02.6 & 17.56 &   1.58 &  64402  &  81 & a & \mathrm{C}\\
100             &     40\ 29.35 , 46\ 22.4 & 19.48 &   2.21 &  67825  & 110 &   & \mathrm{C}\\
101             &     40\ 29.74 , 42\ 31.7 & 19.93 &   1.98 &  63681  & 196 &   & \mathrm{C}\\
102             &     40\ 30.70 , 41\ 46.6 & 19.02 &   2.20 &  67255  &  71 &   & \mathrm{C}\\
103             &     40\ 31.87 , 40\ 45.6 & 18.79 &   1.43 &  69097  & 140 &   & \mathrm{C}\\
\textit{104}    &     40\ 31.94 , 42\ 30.5 & 17.59 &   1.85 &  72697  &  78 & i & \mathrm{C}\\
105             &     40\ 32.09 , 46\ 36.8 & 19.03 &   .... &  64928  &  74 &   & \mathrm{C}\\
106             &     40\ 32.93 , 46\ 45.9 & 17.46 &   .... &  66408  &  51 & k & \mathrm{C}\\
107             &     40\ 33.00 , 45\ 01.1 & 18.55 &   2.27 &  67758  &  81 & k & \mathrm{T}\\
108             &     40\ 33.29 , 43\ 49.1 & 19.98 &   1.77 &  67462  & 113 & k & \mathrm{T}\\
109             &     40\ 33.74 , 40\ 31.2 & 19.60 &   0.49 &  70574  & 175 &   & \mathrm{C}\\
110             &     40\ 34.25 , 39\ 03.9 & 19.43 &   1.48 &  65715  &  97 &   & \mathrm{C}\\
111             &     40\ 35.11 , 44\ 45.1 & 18.70 &   2.10 &  68703  & 159 &   & \mathrm{C}\\
112             &     40\ 35.62 , 41\ 28.8 & 19.24 &   2.41 &  66630  &  94 &   & \mathrm{C}\\
113             &     40\ 36.41 , 44\ 18.3 & 18.77 &   2.11 &  70142  &  78 &   & \mathrm{C}\\
114             &     40\ 36.96 , 42\ 29.0 & 19.90 &   2.03 &  64736  & 205 &   & \mathrm{C}\\
\textit{115}    &     40\ 37.63 , 44\ 30.0 & 19.92 &   2.20 & 122821  & 179 &   & \mathrm{T}\\
\textit{116}    &     40\ 37.66 , 38\ 58.7 & 18.60 &   1.52 &  79042  &  92 & e & \mathrm{C}\\
\textit{117}    &     40\ 37.82 , 45\ 08.8 & 18.10 &   3.09 &  61291  & 156 &   & \mathrm{C}\\
118             &     40\ 37.94 , 42\ 46.8 & 19.36 &   2.16 &  65551  & 136 &   & \mathrm{C}\\
119             &     40\ 38.45 , 43\ 47.9 & 19.62 &   1.42 &  65813  &  78 & a & \mathrm{C}\\
120             &     40\ 38.57 , 42\ 35.4 & 20.35 &   2.27 &  67301  & 170 &   & \mathrm{T}\\
\textit{121}    &     40\ 38.88 , 40\ 20.7 & 18.94 &   1.66 &  57095  & 207 &   & \mathrm{C}\\
122             &     40\ 39.36 , 42\ 52.3 & 18.88 &   2.09 &  66992  &  83 &   & \mathrm{C}\\
123             &     40\ 39.72 , 39\ 07.0 & 17.70 &   2.22 &  67564  & 101 & k & \mathrm{C}\\
124             &     40\ 39.86 , 44\ 45.3 & 19.40 &   2.02 &  64910  &  81 &   & \mathrm{C}\\
125             &     40\ 41.11 , 40\ 10.9 & 19.78 &   2.09 &  66389  & 235 &   & \mathrm{C}\\
126             &     40\ 41.74 , 40\ 23.4 & 19.31 &   2.25 &  67194  & 152 &   & \mathrm{C}\\
127             &     40\ 42.74 , 41\ 30.0 & 19.76 &   1.43 &  68626  & 202 &   & \mathrm{C}\\
128             &     40\ 43.46 , 44\ 44.9 & 20.28 &   2.03 &  65829  & 156 &   & \mathrm{C}\\
129             &     40\ 44.42 , 40\ 19.2 & 19.52 &   2.14 &  67997  & 147 &   & \mathrm{C}\\
130             &     40\ 45.62 , 43\ 26.5 & 20.21 &   1.04 & 65784  & 154 & e & \mathrm{C}\\
\textit{131}    &     40\ 45.86 , 44\ 59.3 & 20.18 &   1.46 & 57661  & 182 & e & \mathrm{C}\\
132             &     40\ 48.00 , 41\ 56.5 & 19.94 &   2.22 & 66938  & 196 &   & \mathrm{C}\\
              \noalign{\smallskip}			    
            \hline					    
            \noalign{\smallskip}			    
            \hline					    
         \end{array}\\
     $$ 
\label{tab1}
\\
 Note. The velocities and nominal cross--correlation errors (in km sec$^{-1}$) for 6 galaxies spectroscopically observed both with the TNG and the CFHT, respectively, are given below:\\
\#38 (106730$\pm$73; 106738$\pm$148); \#53
(68334$\pm$48; 68227$\pm$59); \#65 (67259$\pm$31; 67372$\pm$49); \#98
(66766$\pm$46; 67074$\pm$49); \#107 (67758$\pm$35; 67921$\pm$46); \#108
(67462$\pm$49; 67529$\pm$34).
   \end{table}

\begin{figure*}
\centering
\resizebox{\hsize}{!}{\includegraphics{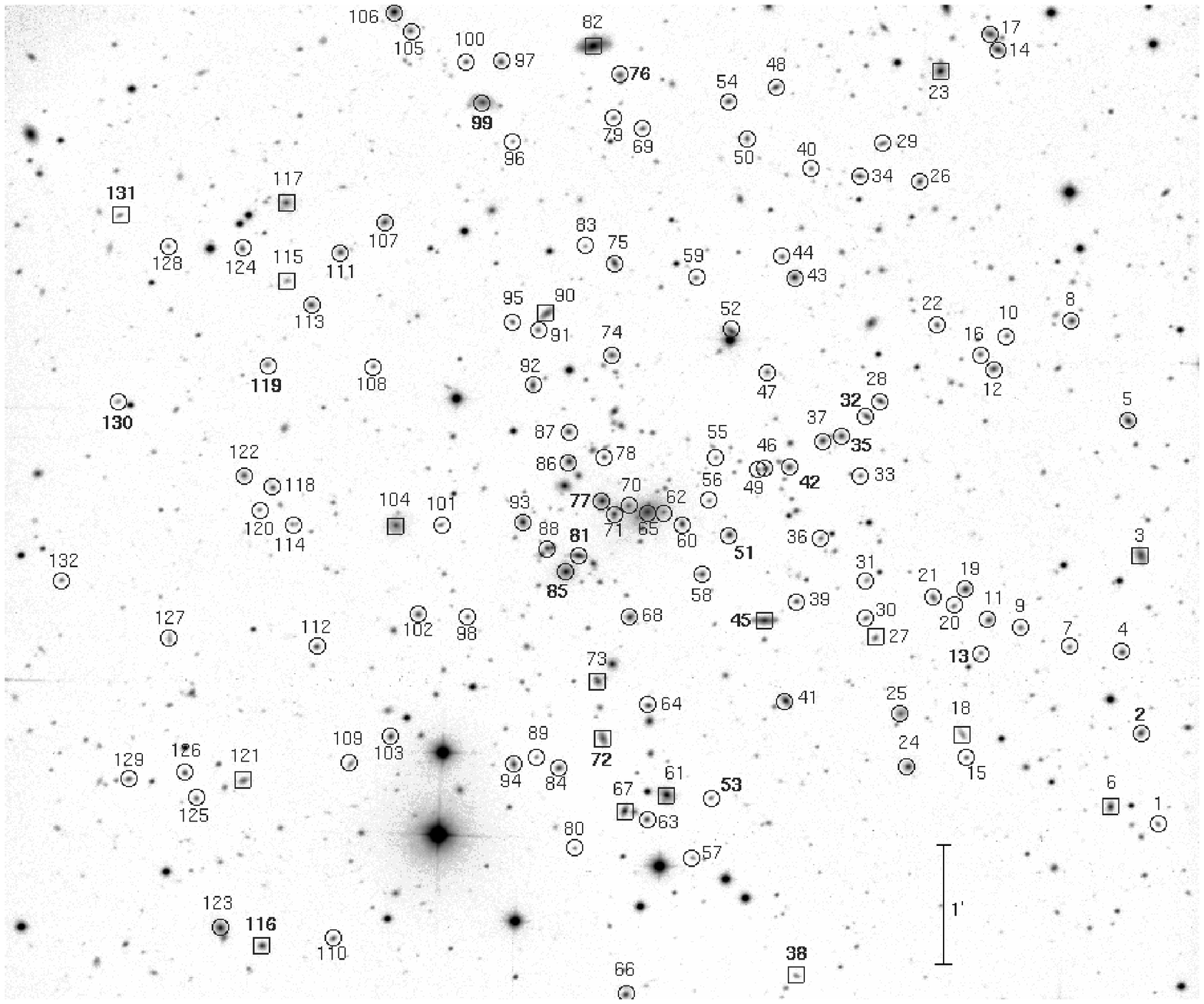}}
\caption{R--band image of A2219 (North at top and East to
left). Galaxies with successful velocity measurements are labeled as
in Table~\ref{tab1}. Circles and boxes indicate cluster member and non
member galaxies, respectively. Labels in bold indicate active galaxies
(see text).}
\label{figimage}
\end{figure*}

Table~\ref{tab1} lists the velocity catalogue\footnote{The
coordinates of non-galaxy objects are available upon request.}
(see also Fig.~\ref{figimage}): identification number of each galaxy,
ID (Column~1); right ascension and declination, $\alpha$ and $\delta$
(J2000, Column~2); R magnitudes (Column~3); B--R colours (Column~4);
heliocentric radial velocities, ${\rm v}=cz_{\sun}$ (in \kss,
Column~5) with assumed errors, $\Delta {\rm v}$, i.e., the nominal
ones given by cross--correlation technique multiplied by 2.3
(Column~6); the code for the spectral classification SC, where ``I''
and ``R'' indicate mid--IR and radio emitting galaxies, respectively
(see Sect.~5), (Column~7); the telescope used to obtain the spectra
(Column~8, T and C indicate TNG and CFHT, respectively).

\section{Optical analysis}

\subsection{Member selection and global properties}

\begin{figure}
\centering
\resizebox{\hsize}{!}{\includegraphics{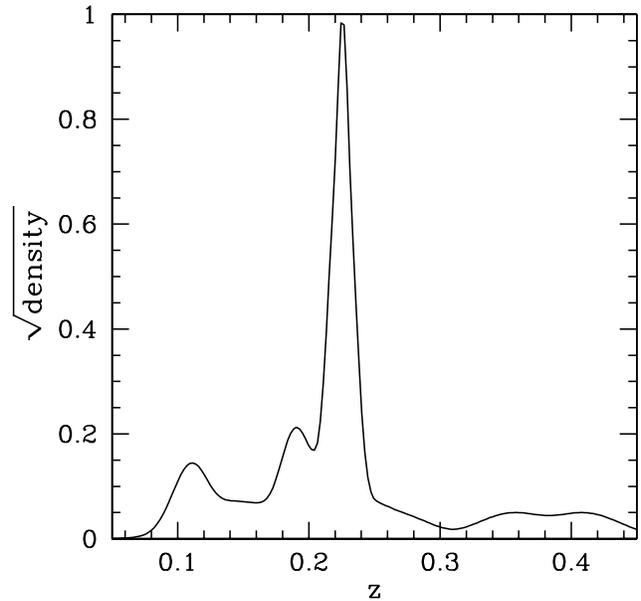}}
\caption{Square root of the redshift galaxy density, as provided by
the adaptive--kernel reconstruction method. Units on the $y$ axis are
arbitrary.}
\label{figden}
\end{figure}

According to the analysis of the redshift distribution based on the
one--dimensional adaptive kernel technique (Pisani \cite{pis93}, see
also Fadda et al.'s \cite{fad96} and Girardi et
al. \cite{gir96}) A2219 appears as a peak at $z\sim 0.225$ in redshift
space (Fig.~\ref{figden}).  Out of 132 galaxies with redshift,
fourteen are foreground objects and three are background objects. In
particular, the foreground galaxies form two peaks, one of six
galaxies at $z\sim 0.11$ and one of eight galaxies at $z\sim 0.19$.

\begin{figure}
\centering
\resizebox{\hsize}{!}{\includegraphics{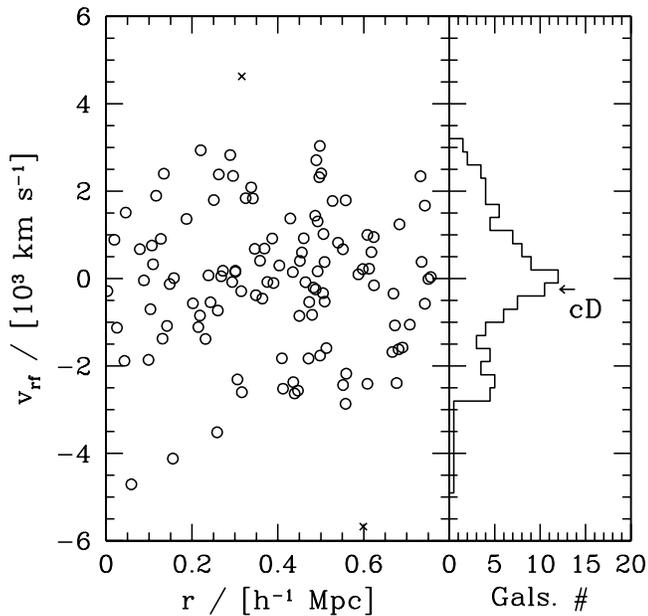}}
\caption{Left panel: rest--frame velocities $\rm{v}_{\rm{rf}}$ vs.
projected cluster--centric distance $r$ of galaxies in the main
peak of Fig.~\ref{figden}. Circles indicate cluster members; crosses
indicate the two galaxies rejected by the ``shifting gapper''
procedure.  Right panel: histogram of rest--frame velocities of
selected 113 cluster members.  The arrow indicates the cD galaxy
velocity.}
\label{figvd}
\end{figure}

In order to reject possible interloper galaxies in the main peak we
apply the procedure of the ``shifting gapper'' of Fadda et
al. (\cite{fad96}), which combines velocity and position
information. This procedure rejects galaxies that are too far in
rest--frame velocity ($\rm{v}_{\rm{rf}}=({\rm v}-\overline{\rm
v})/(1+\overline{\rm v}/c)$) from the main body of galaxies within a
fixed radial bin shifted along the distance from the cluster center.
According to Fadda et al.  prescriptions, we use a gap of $1000$ \ks
and a bin of 0.4 \hh, or large enough to include 15 galaxies. As for
the cluster center, based on the two--dimensional kernel analysis, we
find that the densest peak is centered at
R.A. $16^{\mathrm{h}}\,40^{\mathrm{m}}\,20.56^{\mathrm{s}}$ and
Decl. $+46\degr\,42\arcmin\,42.2\arcsec$, very close to the cD galaxy
position; thus hereafter we assume the cD position as the cluster
center R.A. $16^{\mathrm{h}}\,40^{\mathrm{m}}\,19.87^{\mathrm{s}}$ and
Decl. $+46\degr\,42\arcmin\,41.3\arcsec$. The shifting gapper
procedure rejects two galaxies as non--members. They are
indicated by crosses in Fig.~\ref{figvd}, which shows the velocity
histogram of the 113 selected cluster members. In order to check the
reliability of the member selection procedure, we also apply the
method by den Hartog \& Katgert (\cite{den96}). Out of the initial 132
galaxies with redshift we find the same 113 cluster members.

The mean cluster velocity $\overline{\rm v}=67574\pm136$ \ks
is computed by using the biweight estimator (ROSTAT package; Beers et
al. \cite{bee90}).  We estimate the line--of--sight velocity dispersion,
$\sigma_{\rm v}$, by using the biweight estimator and applying the
cosmological correction and the usual correction for velocity errors
(Danese et al. \cite{dan80}).  We obtain $\sigma_{\rm
v}=1438^{+109}_{-86}$ \kss, where errors are obtained with the
bootstrap technique.  Assuming that the system is in dynamical
equilibrium, the value of $\sigma_{\rm v}$ leads to a value of the
radius of the collapsed, quasi--virialized region of $R_{\rm vir}\sim
R_{200}=\sqrt{3}\,\sigma_{\rm v}/[10{\,\rm H}(z)]\sim 2.2$ \h
(Carlberg et al. \cite{car97}) and a virial mass estimate of
$M(<R_{\rm vir})=2.8^{+0.8}_{-0.7}\times10^{15}$ \msun, (Girardi et
al. \cite{gir98}; Girardi \& Mezzetti \cite{gir01}).

\begin{figure}
\centering
\resizebox{\hsize}{!}{\includegraphics{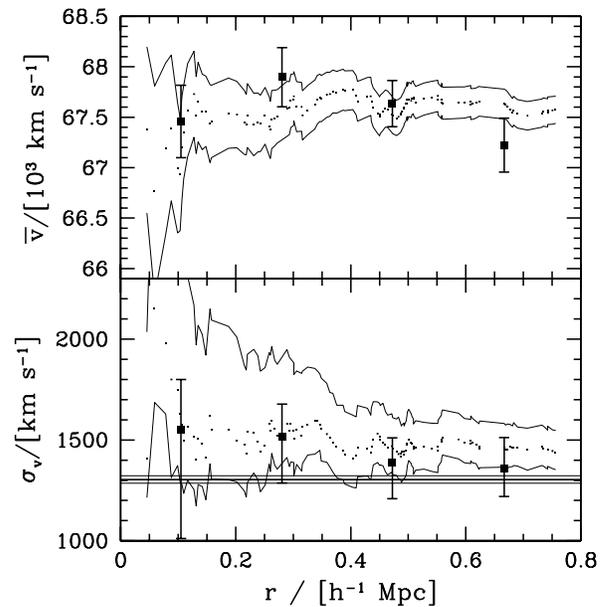}}
\caption{Differential (big squares) and integral (small points) mean velocity
and line--of--sight velocity dispersion profiles. For the differential
profiles, the results for four annuli from the cluster center, each of 0.2
\hh, are shown. For the integral profiles, the mean and dispersion at
a given (projected) radius from the cluster center is estimated by
considering all galaxies within that radius. The error bands
at the $68\%$ c.l. are shown. In the lower panel, the horizontal
lines represent X--ray temperature with the respective errors
transformed in $\sigma_{\rm v}$ (assuming $\beta_{\rm{spec}}=1$ where
$\beta_{\rm{spec}}=\sigma_{\rm v}^2/(kT_{\rm X}/\mu m_{\rm p})$, with $\mu$
the mean molecular weight, $m_{\rm p}$ the proton mass, and $T_{\rm
X}$ the X--ray temperature).}
\label{figprof}
\end{figure}

The high value of $\sigma_{\rm v}$ is confirmed also if we apply a
tighter member selection (i.e., using a bin of 0.25\,\h in the
shifting gapper): only another three galaxies are rejected --those with
the lowest velocities-- resulting in $\sigma_{\rm v}=1355^{+82}_{-82}$
\kss for the remaining galaxies.  Moreover, we compute the
differential and integral mean velocity and velocity dispersion
profiles in Fig.~\ref{figprof}: this analysis shows that the global
values of $\overline{\rm v}$ and $\sigma_{\rm v}$ are already reached
within the central cluster region of 0.1--0.2 \hh.  The conclusion of
the above analyses is that contamination by obvious field interlopers
and/or close cluster companions cannot explain the high value of the
global velocity dispersion.  More probably, this value is connected
with the peculiarity of the internal dynamics of the cluster itself,
which will be analyzed below.

\subsection{Internal dynamics}

As a first step in the study of internal dynamics we analyze the
velocity distribution. We consider three estimators of Gaussianity:
the kurtosis; the skewness, and the scaled tail index (see, e.g.,
Beers et al. \cite{bee91}). We find no significant evidence that the
velocity distribution differs from a Gaussian.  Moreover, we do not
find any evidence of a peculiar velocity of the cD galaxy with respect
to the average velocity of cluster members (see Gebhardt \& Beers
\cite{geb91}).

\begin{figure}
\centering
\resizebox{\hsize}{!}{\includegraphics{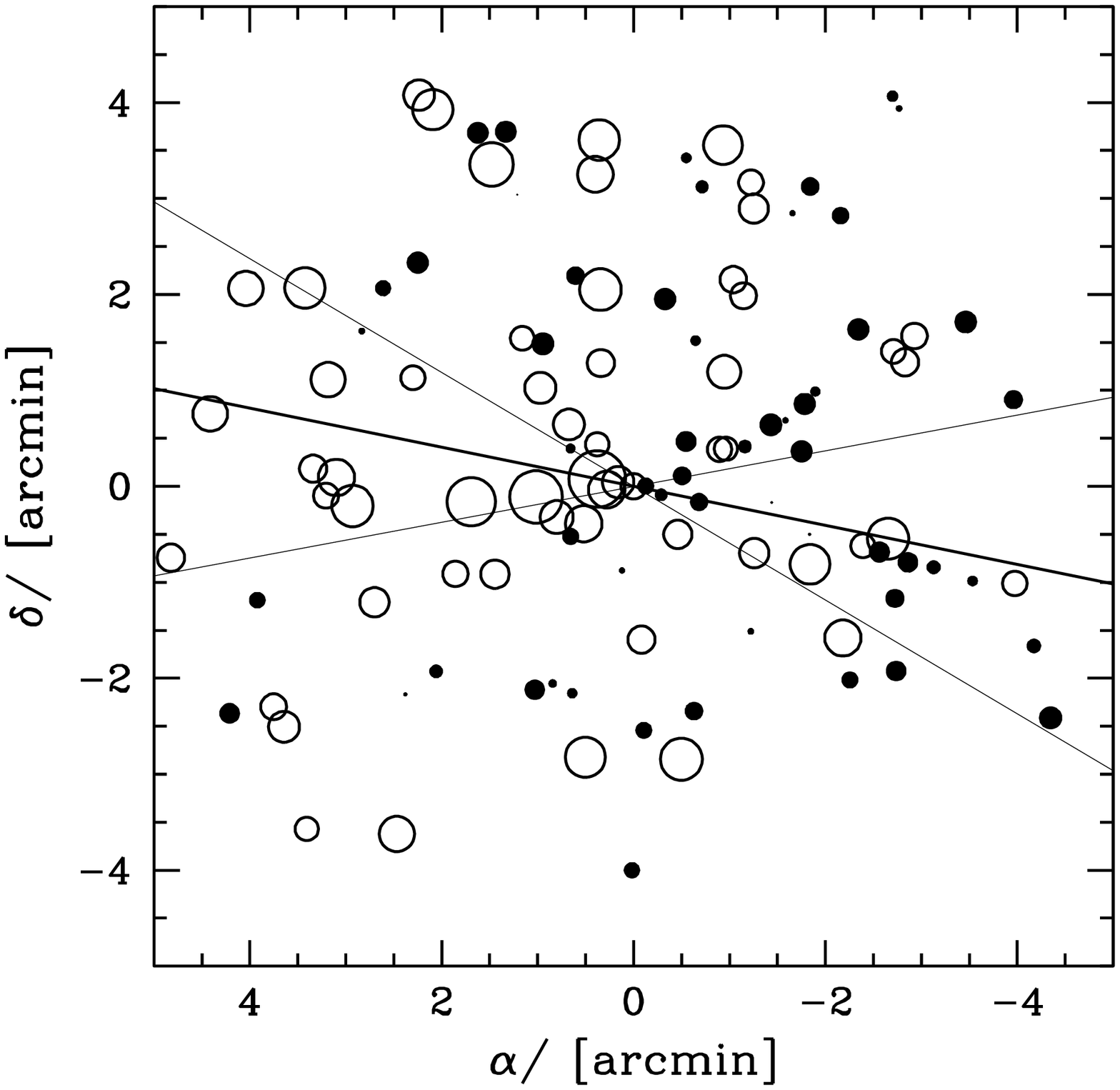}}
\caption{Spatial distribution on the sky of 113 cluster members.  Open and
solid circles indicate low--and high--velocity galaxies: the larger
the symbol, the smaller is the radial velocity.  The plot is centered
on the cluster center (coincident with the cD galaxy).  The solid and
faint lines indicate the position angle of the cluster velocity
gradient and relative errors, respectively.}
\label{figgrad}
\end{figure}

In order to investigate the velocity field we divide galaxies in a low
and a high--velocity samples by using the median cluster velocity and
check the difference between the two distributions of galaxy
positions.  Fig.~\ref{figgrad} shows that low and high--velocity
galaxies are segregated roughly along the E--W direction.  The two
distributions are different at the $99.6\%$ c.l.  according to the
two--dimensional Kolmogorov--Smirnov test (hereafter 2DKS--test; see
Fasano \& Franceschini \cite{fas87}, as implemented by Press et
al. \cite{pre92}). In order to estimate the direction of the velocity
gradient we perform a multiple linear regression fit to the observed
velocities with respect to the galaxy positions in the plane of the
sky (see also den Hartog \& Katgert \cite{den96}; Girardi et
al. \cite{gir96}): we find a PA$=79^{+22}_{-20}$ degrees (measured
counter--clock--wise from the North, see Fig.~\ref{figgrad}).

Moreover, we compute the mean velocity and LOS velocity dispersion
separately for each one of the four samples corresponding to the NE,
SE, SW and NW cluster sectors.  Table~\ref{tab2} shows that mean
velocities are larger in the Western than in the Eastern sectors in
agreement with the above analyses.  As for $\sigma_{\rm v}$, this is
largest in the SE sector and smallest in the NW sector: the difference
is significant at the $98.7\%$ c.l. according to the F--test (Press
et al. \cite{pre92}).

\begin{table}
        \caption[]{Mean galaxy velocities and velocity dispersion (in \kss) in four cluster sectors (see text). N indicates the number of member galaxies in each sector.}
         \label{catalogue}
              $$ 
           \begin{array}{l c c r }
            \hline
            \noalign{\smallskip}
            \hline
            \noalign{\smallskip}

\mathrm{\,\,\,\,\,Sector}      & \mathrm{N} & \overline{\rm v} & \sigma_{\rm v}\,\,\,\,\,\, \\
            \hline
            \noalign{\smallskip}   

  \mathrm{\,\,\,\,\,\,\,NE}  &  30& 66712\pm 272 &1458^{+305}_{-172} \\
  \mathrm{\,\,\,\,\,\,\,SE}  &  26& 67213\pm 330 &1642^{+271}_{-152} \\
  \mathrm{\,\,\,\,\,\,\,SW}  &  25& 68007\pm 288 &1403^{+248}_{-224} \\
  \mathrm{\,\,\,\,\,\,\,NW}  &  31& 68128\pm 187 &1018^{+161}_{-109} \\

              \noalign{\smallskip}			    
            \hline					    
            \noalign{\smallskip}			    
            \hline					    
         \end{array}
     $$ 
\label{tab2}
         \end{table}

\begin{figure}
\centering
\resizebox{\hsize}{!}{\includegraphics{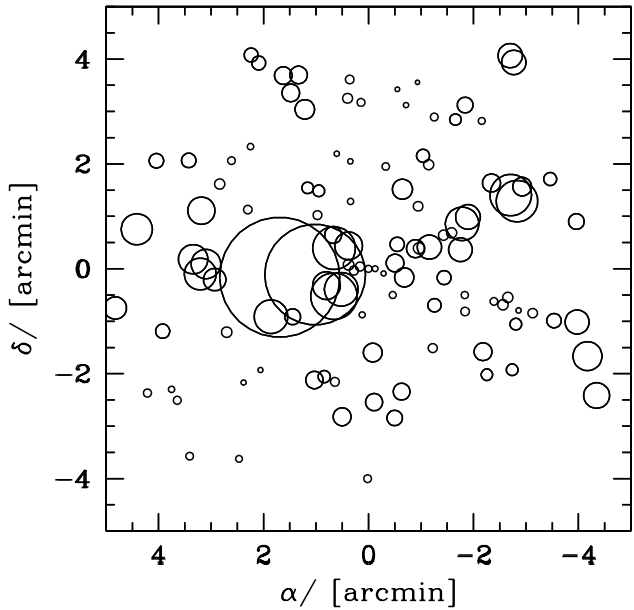}}
\caption{Spatial distribution on the sky of the 113 cluster members,
each marked by a circle: the larger the circle, there is more evidence
for substructure (according to the Dressler \& Schectman test, see
text).  The plot is centered on the cluster center.}
\label{figds}
\end{figure}

To check for the presence of three--dimensional substructure we
compute the $\Delta$--statistics devised by Dressler \& Schectman
(\cite{dre88}) and establish its significance running 1000 Monte Carlo
simulations, in which we randomly shuffle galaxy velocities. The
signal of substructure is significant at the 95\% c.l.. In Fig.~\ref{figds}
we plot the distribution on the sky of all galaxies, each marked by a
circle: the larger the circle, the larger is the deviation
$\delta_{\rm i}$ of the local parameters from the global cluster
parameters, i.e., there is more evidence for substructure.  The most
likely substructure lies close to the cluster center roughly to the SE
of the cD galaxy.

\subsection{Colour segregation and 2-D galaxy distribution}

We check for possible colour segregation of galaxies by using the
110 member galaxies for which we have B--R colours. We
obtain a correlation between $|\mathrm{v}_{\rm{rf}}|$ and colour (at
the $96.3\%$ c.l.).  We divide the sample in two subsamples with
colours bluer and respectively redder than the median B--R
colour=2.12. We obtain that the velocity dispersions of the two
subsamples differ at the $99.5\%$ c.l.  ($\sigma_{\rm
v}=1627^{+144}_{-96}$ and $\sigma_{\rm v}=1100^{+190}_{-160}$ \ks for
the B--R $<$ 2.12 and B--R $>$ 2.12 galaxies, respectively).

\begin{figure}
\centering
\resizebox{\hsize}{!}{\includegraphics{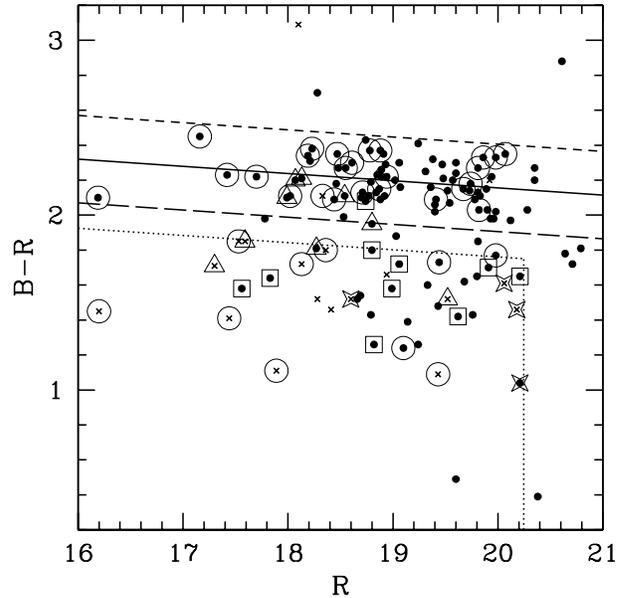}}
\caption{B--R vs. R diagram for galaxies with available spectroscopy:
small dots and crosses denote cluster and field members, respectively.
Large symbols denotes spectroscopic classified galaxies: circles,
triangles, squares, and stars indicate $k$--, $i$--, $a$--, and
$e$--galaxies, respectively.  The solid line gives the best--fit CMR;
the dashed lines are drawn at $\pm$0.25 mag from the CMR. According to
our working definition in Sect.~3.3 cluster members are divided in:
{\em blue} and {\em red} galaxies (below and above the long--dashed
line, respectively); {\em very red} galaxies lie above the solid
line. The left--down region delimited by the dotted lines defines the
locus of Butcher--Oemler galaxies (Sect.~5).}
\label{figcm}
\end{figure}

To further investigate the above difference we also use the
colour--magnitude relation (hereafter CMR), which indicates the
early--type galaxy locus,
\begin{equation}
\rm{B}-\rm{R}=2.976-0.041\cdot \rm{R},
\end{equation}
obtained by using a 2 sigma--clipping fitting procedure (see
Fig.~\ref{figcm}). {\em Blue} objects are defined to be those galaxies
at least 0.25 mag bluer in B--R than the colour of the CMR. These {\em
blue} galaxies have typically B--R$<2$, thus include Sa spiral
galaxies and exclude elliptical galaxies (according to the expected
typical colours).  The remaining objects, which we define the {\em
red} sample, should consist most of ellipticals and lenticulars.
Moreover, we further divide {\em red} galaxies in two subsamples: {\em
very red} and {\em not--very red} galaxies, depending on their
location above or below the CMR.  Table~\ref{tab3} shows
$\overline{\rm v}$ and $\sigma_{\rm v}$ estimates for {\em blue} and
{\em red} galaxies, and separately for {\em very} and {\em not--very
red} galaxies. We point out that the populations of {\em very red} and
{\em not--very red} galaxies are different. In fact, the $\sigma_{\rm
v}$ of {\em not--very red} galaxies is larger than the $\sigma_{\rm
v}$ of {\em very red} ones (at the $99.5\%$ c.l. according to the
F--test). Moreover, the mean velocity of {\em not--very red} galaxies
is smaller than that of {\em very red} ones (at the $99.5\%$
c.l. according to the means--test).

\begin{table}
        \caption[]{Mean velocities and velocity dispersions (in \kss) for galaxies with different colors (see text). N is the number of galaxies of each type.}
         \label{catalogue}
              $$ 
           \begin{array}{l c c r }
            \hline
            \noalign{\smallskip}
            \hline
            \noalign{\smallskip}

\mathrm{\,\,\,\,\,\,\,\,\,\,\,\,\,\,\,\,\,\,\,Population}      & \mathrm{N} & \overline{\rm v} & \sigma_{\rm v}\,\,\,\,\,\,\, \\  
            \hline
            \noalign{\smallskip}   

  blue\mathrm{\,\,galaxies}                 &   33 & 67676\pm 255 & 1439^{+177}_{-139} \\
  red\mathrm{\,\, galaxies}                 &   77 & 67568\pm 163 & 1423^{+146}_{-123} \\
  (not\,\, very)\,\,red{\bf \mathrm{\,\, galaxies}}&   40 & 66942\pm 265 & 1655^{+215}_{-168} \\
  (very)\,\,red{\bf \mathrm{\,\, galaxies}}&   37 & 67969\pm 173 & 1032^{+172}_{-126} \\

              \noalign{\smallskip}			    
            \hline					    
            \noalign{\smallskip}			    
            \hline					    
         \end{array}
     $$ 
\label{tab3}
         \end{table}

\begin{figure}
\centering
\includegraphics[width=7cm]{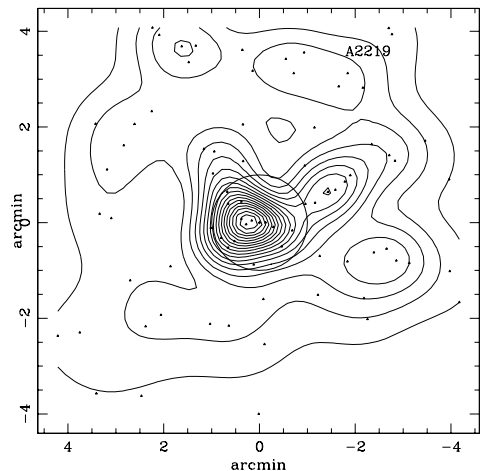}
\caption{Spatial distribution on the sky and relative isodensity
contour map of 80 bright cluster (R$<19.7$ galaxies), obtained with
the adaptive--kernel method (cf. Pisani \cite{pis93},\cite{pis96}).
The plots are centered on the cluster center.  The 1$\arcmin$ circle
is centered on the cD galaxy.}
\label{figxy}
\end{figure}

We also analyze the two--dimensional spatial distribution of galaxies.
First, we consider bright cluster members, i.e. the sample of 80
galaxies with R$<19.7$ for which we have a good level of spatial
completeness (see Sect.~2).  Fig.~\ref{figxy} shows the presence of
two clumps in the cluster center which defines a SE--NW elongated
structure (PA$\sim 110 \degr$): the main peak roughly coincides with
the cD position, and the secondary peak -- at the West -- traces the
filament of high velocity galaxies well visible in Fig.~\ref{figimage}
(from no. 49 to no. 28). The cD also shows an elongation in the SE--NW
direction described by the position angle on the celestial sphere
PA$=114 \pm 3$ degrees (Smail et al. \cite{sma95}).

\begin{figure}
\centering
\includegraphics[width=7cm]{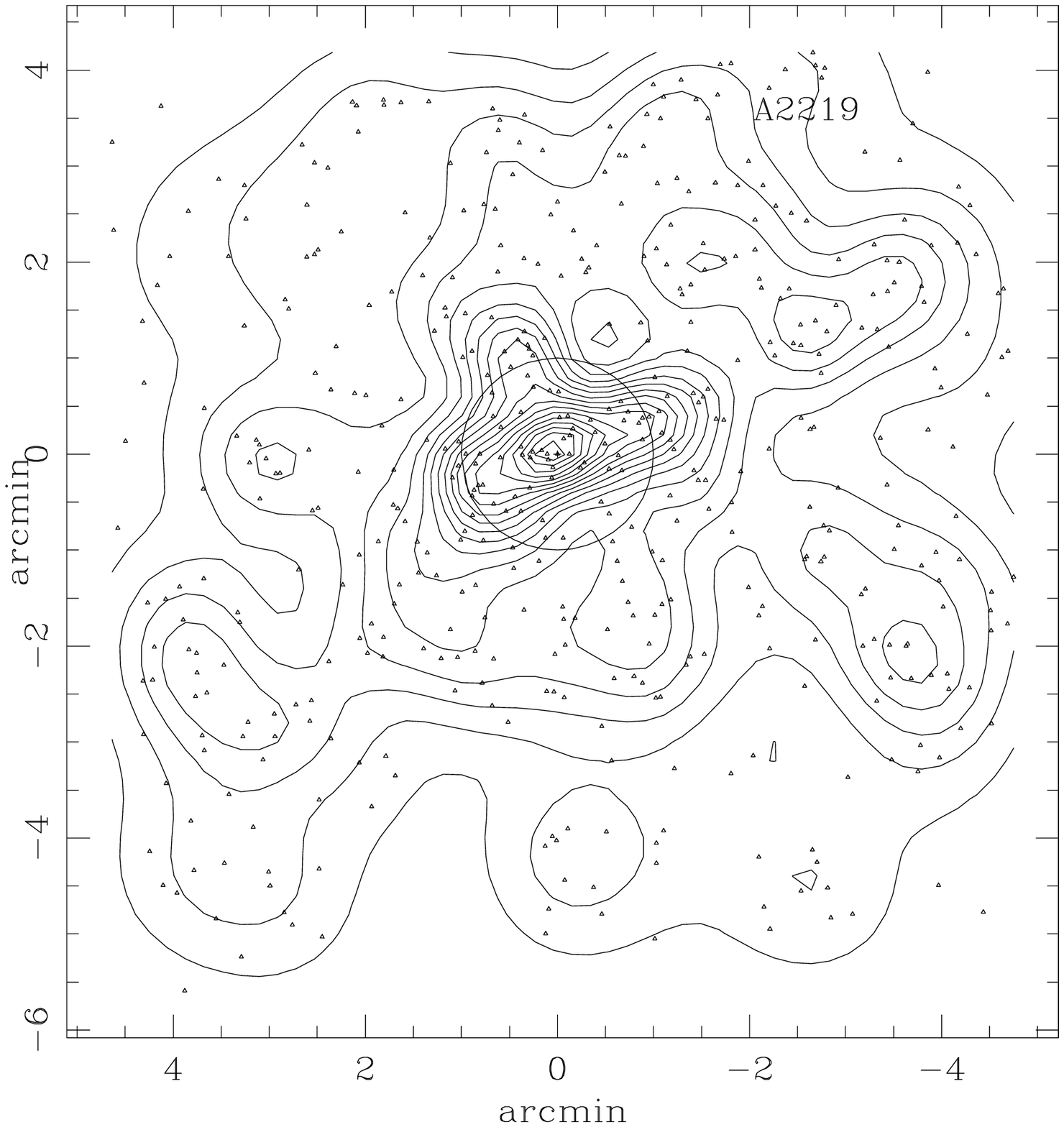}
\includegraphics[width=7cm]{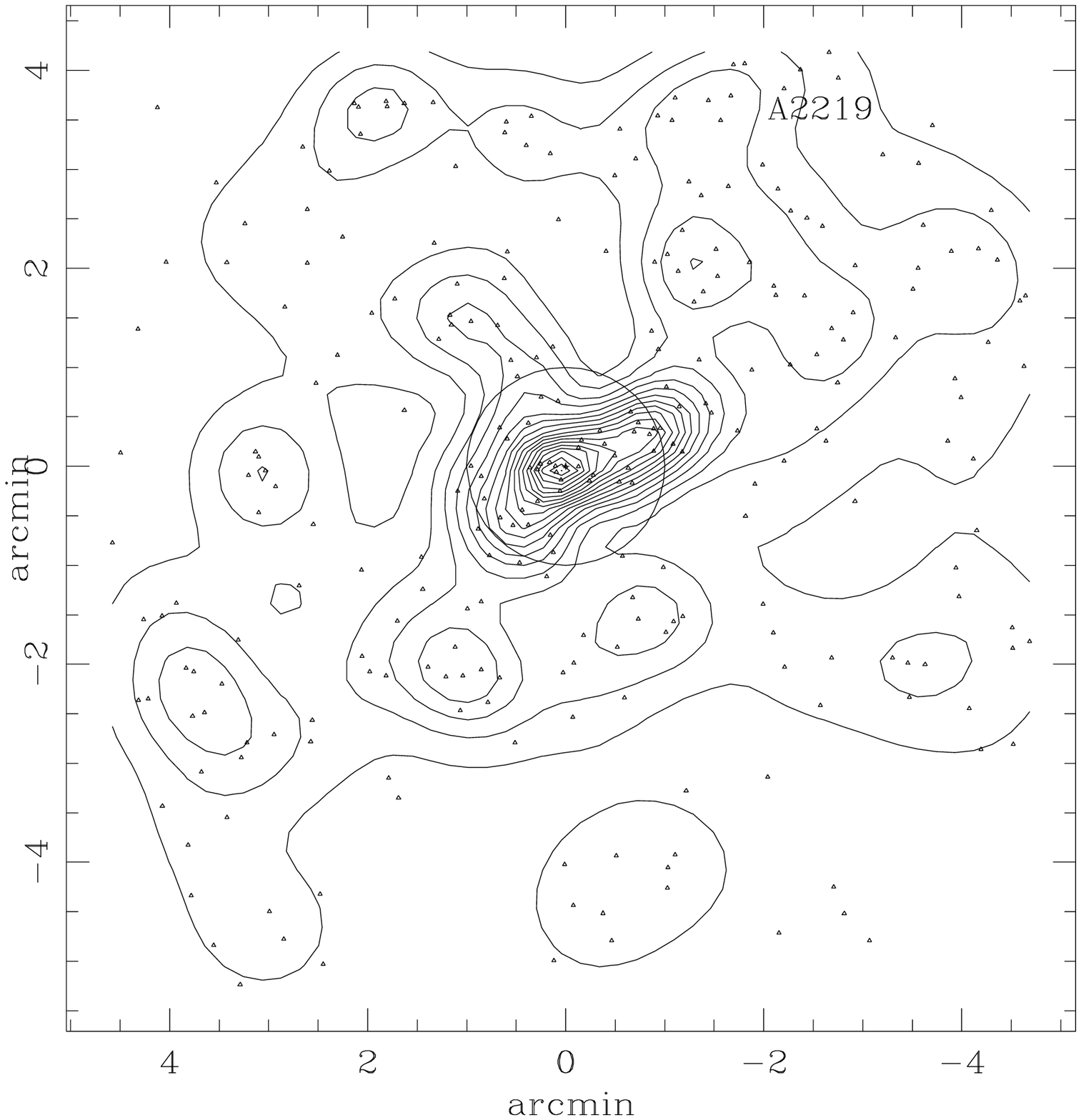}
\includegraphics[width=7cm]{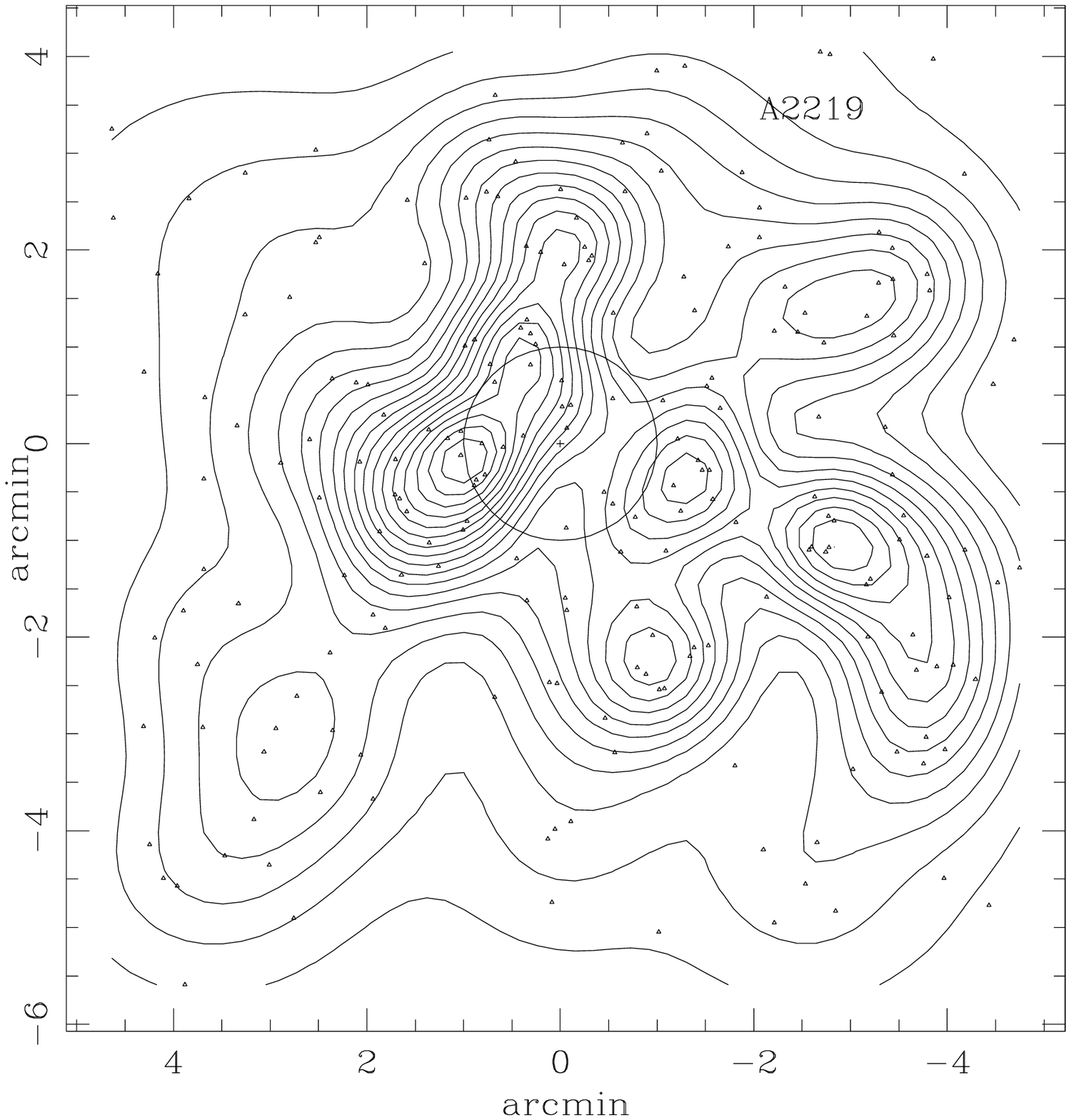}
\caption{Same as Fig.~\ref{figxy} but for likely cluster members
selected on the basis of the B--I CMR in the photometric sample of
Smail et al. (\cite{sma98}). Top panel: the 484 I$<$22
galaxies. Middle and bottom panels: two subsamples corresponding to
very red and not very red galaxies (273 and 211 galaxies,
respectively).}
\label{figxy3}
\end{figure}

Then, to study fainter galaxies and to work with a much more robust
statistics, we consider the B and I photometric data of Smail et
al. (\cite{sma98}) and define likely members using the
CMR. Specifically, we select 484 likely cluster members considering
objects (I$\,<22$, stellar index $<0.9$) within 0.25 mag from the
$\rm{B}-\rm{I}=5.398-0.115\cdot \rm{I}$. The top panel of
Fig.~\ref{figxy3} shows a main structure centered on the cD galaxy and
elongated in the SE--NW direction, with a PA of $\sim 110\degr$ for
the central region (along the filament described above) and a larger
PA for the external region.  A similar structure is also traced by the
subsample of very red galaxies -- those above the CMR
(Fig.~\ref{figxy3} middle panel). Instead, the distribution of not
very red galaxies shows three main clumps aligned in the NE--SW
direction (Fig.~\ref{figxy3} bottom panel). In particular, the densest
peak is located at about $1\arcmin$ South--East of the cD, in the
position of the substructure suggested by the Dressler--Schectman test
(Sect.~3.2).

\section{X--ray observation and analysis} 

The X--ray analysis of A2219 is performed on the archival data of the
Chandra ACIS--S observation 800072 (exposure ID \#896, P.I.:
J. Houck). The pointing has an exposure time of 42.8 ks. Data
reduction is performed by using the package CIAO\footnote{CIAO is
freely available at http://asc.harvard.edu/ciao/} (Chandra Interactive
Analysis of Observations) on chip S3 (field of view $\sim
8.5\arcmin\times 8.5\arcmin$). First, we remove events from the level
2 event list with a status not equal to zero and with grades one, five
and seven. Then, we select all events with energy between 0.3 and 10
keV. In addition, we clean bad offsets and examine the data filtering
out bad columns and removing times when the count rate exceeds three
standard deviations from the mean count rate per 3.3 s interval. We
then clean S3 chip for flickering pixels, i.e., times where a pixel
has events in two sequential 3.3 s intervals. The resulting exposure
time for the reduced data is 42.3 ks.

\begin{figure*}[!ht]
\centering
\resizebox{\hsize}{!}{\includegraphics{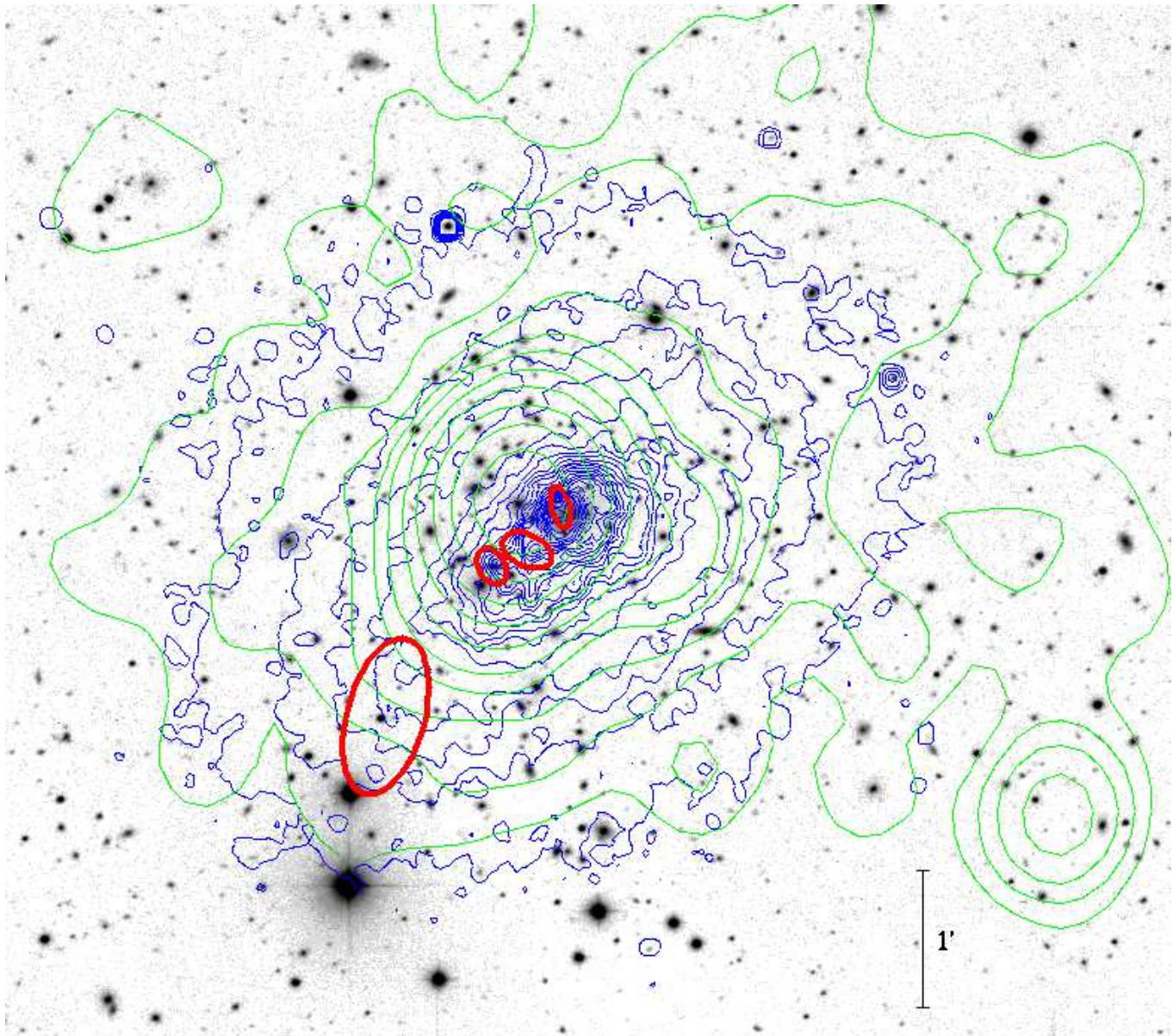}} 
\caption{R--band image of the cluster A2219 with, superimposed, the
contour levels of the Chandra image (blue) and of a low--resolution
VLA image (green) by Bacchi et al. (\cite{bac03}). The red ellipses
represent the structures detected by Wavdetect. North is at top and
East to left.}
\label{figisofote} 
\end{figure*}

In Fig.~\ref{figisofote} we plot an R--band image of the cluster with
superimposed the X--ray contour levels of the Chandra image and the
radio contour levels of a low--resolution VLA image by Bacchi et al.
(\cite{bac03}). The elongated shape of the cluster is clearly
visible. By using the CIAO package Sherpa we fit an elliptical Beta
model to the X--ray photon distribution to quantify the departure from
the spherical shape. The model is defined as follows:
\begin{equation}
f(x,y)=f(r)=A/[1+(r/r_0)^2]^{\alpha}
\end{equation}
where the radial coordinate $r$ is defined as
$r(x,y)=[X^2(1-\epsilon)^2+Y^2]^{1/2}/(1-\epsilon)$,
$X=(x-x_0)\,\cos\,\theta+(y-y_0)\,\sin\,\theta$ and
$Y=(y-y_0)\,\cos\,\theta-(x-x_0)\,\sin\,\theta$. Here $x$ and $y$ are
physical pixel coordinates on chip S3. The best fit centroid position
is located 12$\arcsec$ South--East of the dominant galaxy. The best
fit core radius, the ellipticity and the position angle are
$r_0=69.7\pm$1.3 arcsec (177$\pm$3 \kpc), $\epsilon=0.339\pm$0.006 and
PA=128.8$\pm$0.6 degrees, respectively. The elongated shape of
the cluster is also in agreement with Rizza et al's (\cite{riz98})
result, who found evidence of elongations and excesses in the X-ray
emission relative to a circular model on ROSAT data (see their Table
7).

\begin{figure}[!ht]
\centering
\resizebox{\hsize}{!}{\includegraphics{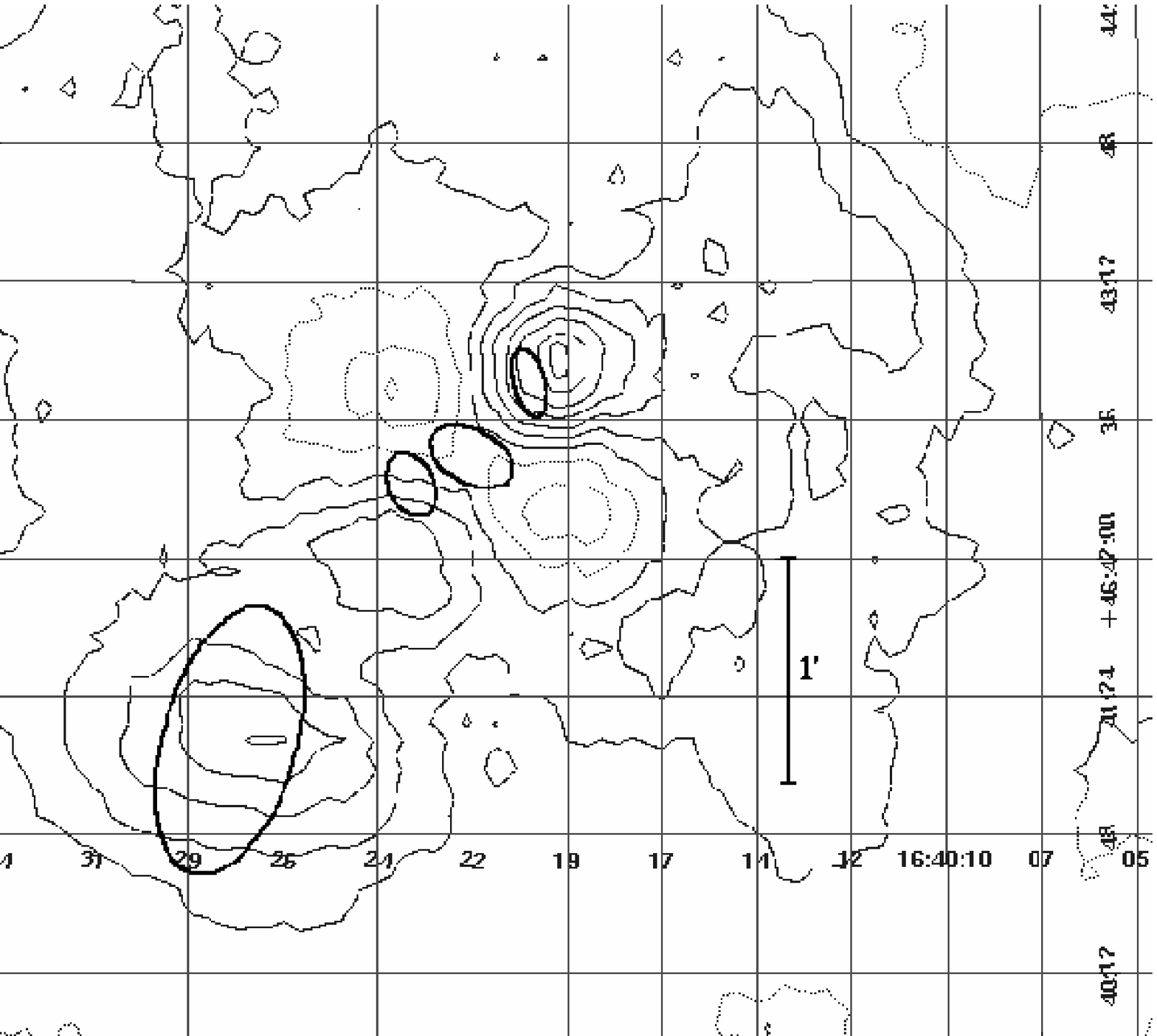}}
\caption{Residual map of the elliptical Beta model fit (see text). Solid and
dotted contour lines indicate regions with positive and negative
residuals, respectively. Ellipses represent the structures detected by
Wavdetect.}
\label{figskew}
\end{figure}

In order to detect possible substructures in A2219 we perform a
wavelet analysis by running the task CIAO/Wavdetect on the chip
S3. The task was run on different scales in order to search for
substructures with different sizes. The significance
threshold\footnote{see \S~11.2 of the CIAO Detect Manual (software
release version 2.2.1, available at the WWW site
http://cxc.harvard.edu/ciao/manuals.html)} was set at $10^{-6}$.  The
results are shown in Fig.~\ref{figisofote}. Thick ellipses represent
four significant substructures found by Wavdetect. The principal one,
located at R.A. $16^{\mathrm{h}}\,40^{\mathrm{m}}\,20.1^{\mathrm{s}}$
and Decl. $+46\degr\,42\arcmin\,45\arcsec$, is centered on the cD
galaxy. There are two more significant peaks in the core of the
cluster located at
R.A. $16^{\mathrm{h}}\,40^{\mathrm{m}}\,23.1^{\mathrm{s}}$ and
Decl. $+46\degr\,42\arcmin\,19\arcsec$, and at
R.A. $16^{\mathrm{h}}\,40^{\mathrm{m}}\,21.6^{\mathrm{s}}$ and
Decl. $+46\degr\,42\arcmin\,26\arcsec$, respectively.
Finally, outside the core, we find a fourth structure at
R.A. $16^{\mathrm{h}}\,40^{\mathrm{m}}\,27.7^{\mathrm{s}}$ and
Decl. $+46\degr\,41\arcmin\,13\arcsec$, about 2$\arcmin$ South--East of
the dominant galaxy. The presence of a photon excess in this position
is also evident from Fig.~\ref{figskew}, where we plot the residual
map of the elliptical Beta model fit. A peak of positive residuals is
observed at the position of the fourth structure.

Passing to the spectral analysis of the cluster X--ray photons, we
first compute a global estimate of the ICM temperature. The
temperature is computed from the spectrum of the cluster within
a circular aperture of 3$\arcmin$ radius around the cluster
center. Fixing the absorbing galactic hydrogen column density at
1.75$\times$10$^{20}$ cm$^{-2}$, computed from the HI maps by Dickey
\& Lockman (\cite{dic90}), we fit a Raymond--Smith (\cite{ray77})
spectrum using the CIAO package Sherpa with a $\chi^{2}$
statistics. We find a best fitting temperature of
$T_{\rm X}=\,$10.3\,$\pm\,0.3$ keV, in agreement with previous estimates
by Allen (\cite{all00}) and White (\cite{whi00}).

\begin{figure}
\centering \resizebox{\hsize}{!}{\includegraphics{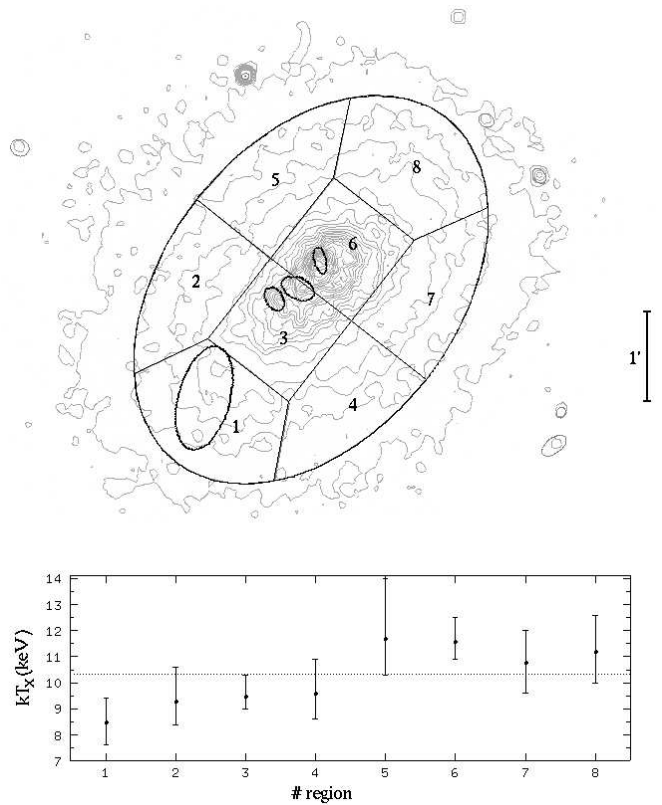}}
\caption{Temperature map of A2219. The cluster area is divided in
eight numbered sectors (top panel). Sectors are superimposed to the
contour levels of the Chandra image. Ellipses show the four structures
detected in the X--ray image. The graph in the bottom panel shows the
temperature computed in each sector.}
\label{figtempmap}
\end{figure}

\begin{figure}
\centering
\resizebox{\hsize}{!}{\includegraphics{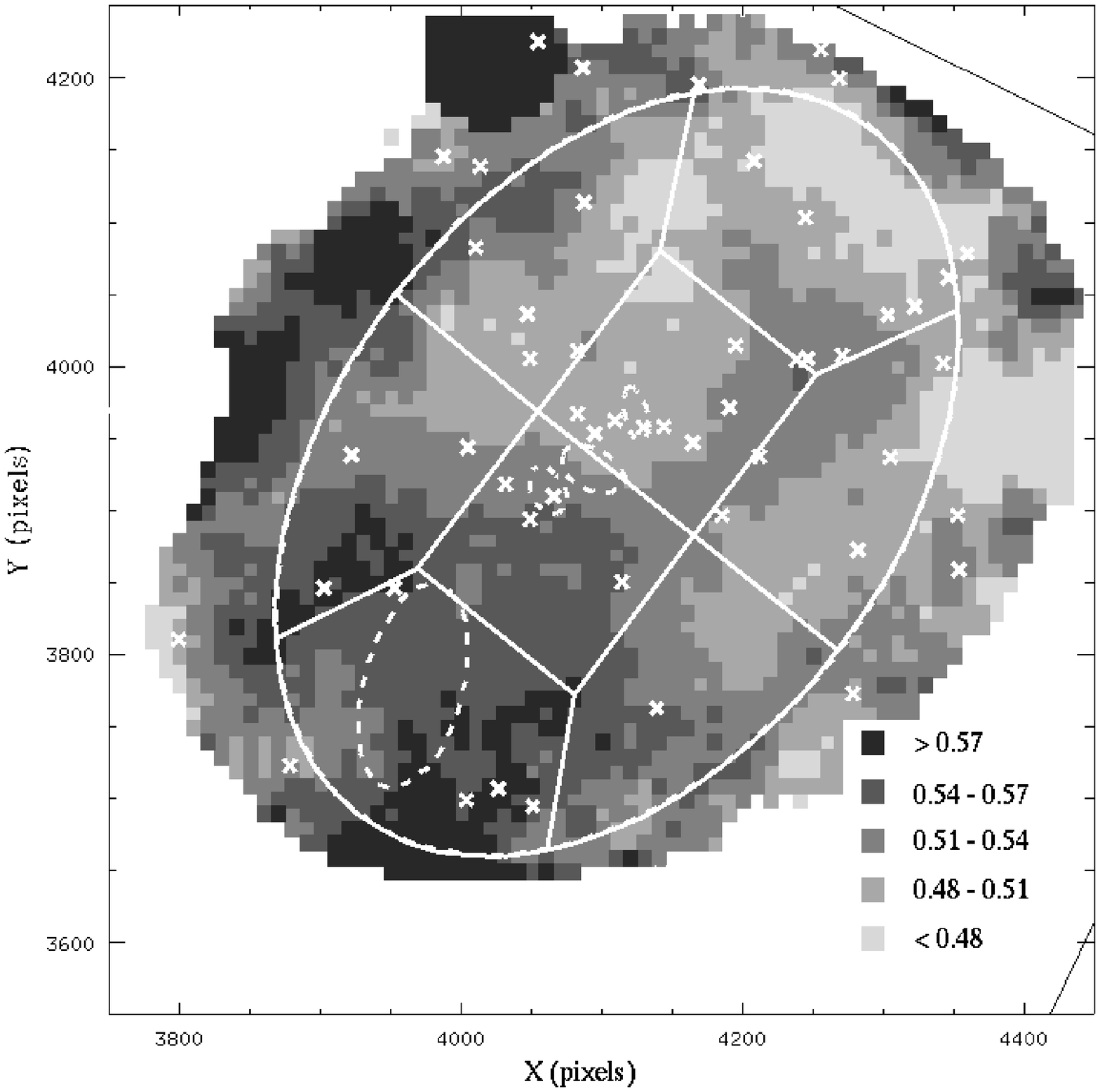}}
\caption{Softness ratio map of A2219 (see text). Superimposed to the
map are the eight sectors (see Fig.~\ref{figtempmap}). Dashed ellipses
are the structures detected in the X--ray image and crosses show the
positions of the cluster members.}
\label{figsoftness}
\end{figure}

To detect the possible presence of temperature gradients in the
cluster we divide the cluster in eight sectors, as shown in
Fig.~\ref{figtempmap}, and compute the temperature in each of them.
We find a temperature gradient from the cooler SE sectors to the
hotter NW ones. This pattern is also confirmed by a softness ratio
analysis of the Chandra image. We define the softness ratio as
$SR\equiv(S-H)/(S+H)$, where $S$ is the number count of photons with
soft energies in the range 0.5--2 keV, while $H$ is the number count
of photons with hard energies in the range 2--7 keV. Then we count
soft and hard photons in circles with 50 pixels aperture radius
sliding on a 200$\times$200 pixels grid. Background photon counts are
subtracted by using an ACIS background event file by
M. Markevitch. The background subtracted values of the softness ratios
are plotted in Fig.~\ref{figsoftness}. In this figure different grey
levels identify regions with different $SR$s. High $SR$ (lower
temperature) regions are concentrated mainly in sectors 1, 2, 3 and 4,
in agreement with the temperature map shown in Fig.~\ref{figtempmap}.

\section{Active galaxies}

The star formation properties of the galaxies in our sample are
examined on the basis of the spectral classification introduced in
Sect.~2 (see Table~\ref{tab1}).  Out of 60 classified galaxies, 46 are
member galaxies: 29/37 $k$--galaxies, 6/9 $i$--galaxies, 1/4
$e$--galaxy, and 10/10 $a$--galaxies.  The relative fraction of
cluster $a$--galaxies is $22\%$. For different EW(H$\delta$) levels we
find $13\%$ and $37\%$ ($>5$ and $>3$ \AA, respectively). For
comparison, Poggianti et al. (\cite{pog99}) found a $21\%$ of k+a/a+k
galaxies with EW(H$\delta$)$>3$ \AA$\,\,$ and Balogh et
al. (\cite{bal99}) found a fraction a $\lesssim 4\%$ of $K+A$ with
EW(H$\delta$)$>$5 \AA. An open question is whether or not strong
H$\delta$ galaxies are typical of cluster environments (cf., e.g.,
Poggianti et al. \cite{pog99} with Balogh et al. \cite{bal99}) since
these galaxies could be produced by a cluster merger (Bekki
\cite{bek99}). Very interestingly, we find 0/14 $a$--galaxies in the
field against 10/46 in cluster environment.

\begin{figure}
\centering
\resizebox{\hsize}{!}{\includegraphics{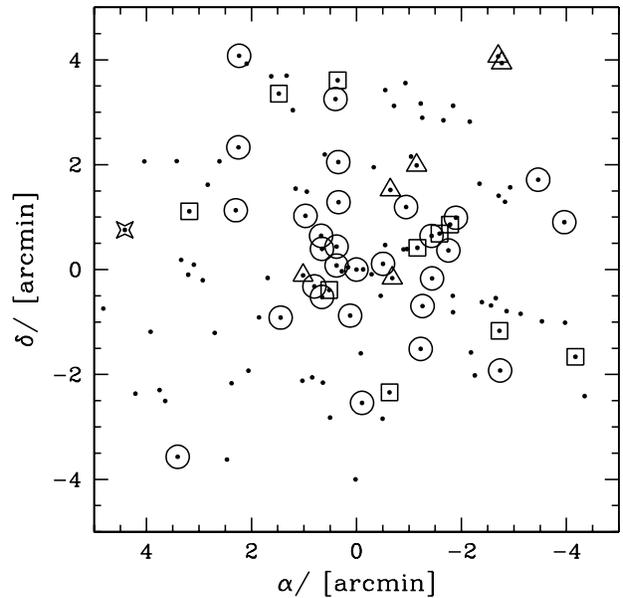}}
\caption{Spatial distribution on the sky of 113 cluster members.
Large symbols denotes spectroscopic classified galaxies: circles,
triangles, squares, and stars indicate $k$--, $i$--, $a$--, and
$e$--galaxies, respectively.}
\label{figspectra}
\end{figure}

We check for possible spectral segregations of cluster galaxies, both
in position and in velocity space by comparing the dynamics of passive
$k$--galaxies with that of active $e$-- plus $a$--galaxies.  We find a
larger velocity dispersion in the case of active galaxies, but the
difference is not significant.  Moreover, we verify that the two
subsamples do not differ in the distribution of galaxy positions by
using the 2DKS--test. Fig.~\ref{figspectra} shows the distribution of
spectroscopically--classified galaxies: 4/25 and 7/21 are classified
as active in the central and external cluster regions (within and
outside 2$\arcmin$).  As for central active galaxies, three are located
in the NW filament of high velocity galaxies in the central region,
corresponding to the NW subclump identified in the two--dimensional
analysis of Sect.~3.3. The fourth galaxy (no. 81) lies at the SE of
the cD galaxy.

As for the interpretation of spectra, the only detected $e$--galaxy is
very blue (the bluest within this sample) as generally found for
starburst models.  Nine of ten $a$--galaxies are classified as {\em
blue} according to our definition (see Fig.~\ref{figcm}). Thus their
spectra can be explained by ``post--starburst'' models which reproduce
strong H$\delta$ EW and no significant emission lines in a quiescent
phase soon after a starburst (about between a few Myr and 1.5 Gyr;
e.g., Poggianti et al. \cite{pog99}). Alternatively, strong H$\delta$
EW could be obtained during a period of $1-2$ Gyr which follows the
halting of the star formation after a 1 Gyr period of constant star
formation (Morris et al. \cite{mor98}).

Galaxy no. 81 shows the strongest H$\delta$ EW, and it is therefore
spectroscopically classified as 'a'. However, it also shows evidence
of mild [OII] emission, EW $\simeq 3$~\AA. It would have been
classified 'e(a)' in the scheme of Poggianti et al. (\cite{pog99}). The galaxy
coincides with one of the X--ray peaks, identified at the
smallest wavelet scale.  The X--ray emission could therefore be
point--like, and consistent with emission from an AGN. The lack of
strong emission lines could be explained if there is substantial dust
obscuration.  This interpretation is consistent with the fact that
e(a) galaxies are generally associated with dusty star--forming
galaxies (Poggianti et al. \cite{pog99}).  The AGN emission should then be
re--emitted in the infrared (IR).  Although galaxy no. 81 was not
detected at 15 $\mu$m with ISOCAM by Barvainis et al. (\cite{bar99}), one can
see in Fig.~4 of Barvainis et al. (\cite{bar99}) low--level isocontours of
mid--IR emission centered on this galaxy. Given that their rms noise
level is $\approx 0.11$ mJy, and their faintest detected source has a
15 $\mu$m flux density of 0.53 mJy, the 15 $\mu$m flux density of
galaxy 81 should be $\la 0.4$--0.5 mJy.

Of the 5 mid--IR sources detected by Barvainis et al. (\cite{bar99}) in the
field of A2219, three are in our spectroscopic sample. These are the
galaxies no. 42, 45 and 72 (respectively no. 1, 2, and 3 in Barvainis
et al.'s Table 1).  Only galaxy no. 42 is a cluster member. The other
two are foreground galaxies. Galaxy no. 42 is an a--type galaxy, with
an H$\delta$ EW $\simeq 8$~\AA. Its spectrum shows evidence of a mild
[OII] emission, EW $\simeq 3$~\AA. It would have been classified
'e(a)' in the scheme of Poggianti et al. (\cite{pog99}). The flux in the [OII]
line can be used to compute the star formation rate of the galaxy,
using the relation of Kennicutt (\cite{ken98}). We find SFR~$\simeq 2 \pm 1 \,
h^{-2} \, M_{\odot}$~yr$^{-1}$.

\begin{figure}
\centering
\resizebox{\hsize}{!}{\includegraphics{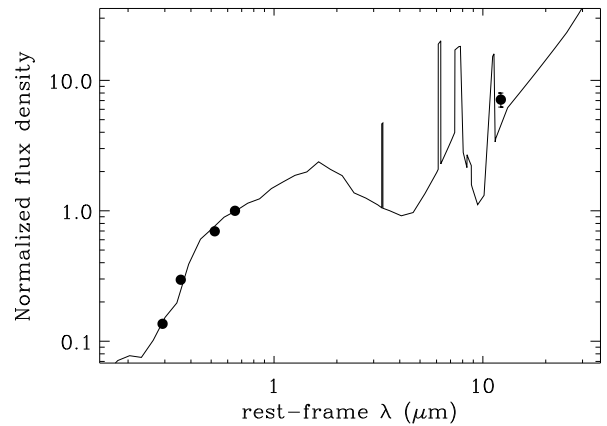}}
\caption{Optical--mid IR spectral energy distribution of galaxy
no. 42. The best--fit GRASIL model is also shown (solid tree).}
\label{figbiv}
\end{figure}

We have examined the optical--mid--IR spectral energy distribution
(SED) of galaxy no. 42, by comparing the observed data (taken from
Barvainis et al. \cite{bar99}) with the GRASIL models of Silva et
al. (\cite{sil98}). The GRASIL model takes into account the effects of several
kinds of dust particles on the reprocessing and obscuration of the
stellar radiation (see:
http://web.pd.astro.it/granato/grasil/grasil.html).  We find that the
observed SED of galaxy no. 42 is best--fitted by a model of a young (3
Gyr--old) early--type spiral, with significant ongoing star formation
(see Fig.~\ref{figbiv}). The best--fit model SED is used to compute the galaxy
total IR luminosity, $\simeq 4 \, 10^{10} \, h^{-2} \, L_{\odot}$.
The galaxy star formation rate, SFR, is derived from the total IR
luminosity using the relation of Kennicutt (\cite{ken98}). We obtain
SFR~$\simeq 7.5 \pm 0.9 \, h^{-2} \, M_{\odot}$~yr$^{-1}$. This value
is about 4 times higher than the estimate we obtain from the [OII]
line flux. This is expected because of significant dust extinction at
short wavelengths.  As a matter of fact, from the best--fit model SED
we estimate that dust extinction reduces the flux in the [OII] line by
a factor $\sim 3$.

We conclude that galaxy no. 42 is a dusty star--forming galaxy, rather
than a post--starburst galaxy. We thus find that both e(a) galaxies of
our sample are dusty, active galaxies, a finding that supports the
interpretation given by Poggianti et al. (\cite{pog99}) for these kind of
galaxies.

With our spectral classification of a subset of A2219 galaxies, we can
try to shed light on the nature of the 'blue' galaxies which are
responsible for the Butcher--Oemler (BO, hereafter) effect (Butcher \&
Oemler \cite{but78}).  We have translated the original Butcher \&
Oemler's definition of 'blue' galaxies to the cluster redshift and the
B and R filters, by using the tables of Fukugita et
al. (\cite{fuk95}), and taking into account the Galactic absorption in
the direction of A2219 (as given in the NASA Extragalactic
Database). The dotted lines in Fig.~\ref{figcm} delimit the region of
the colour--magnitude diagram where the BO galaxies are located. One
can see that the BO population is of a mixed nature. First, there is a
substantial amount of field galaxies, of all spectroscopic
types. Among BO cluster members, the dominant population is that of
a--type galaxies. As we have shown, two of these a--type galaxies are in
fact dusty active galaxies, rather than pure post--starburst
objects. We thus conclude that the BO population of blue galaxies in
the cluster A2219 is mostly composed of galaxies having experienced a
recent starburst. In some of these galaxies the star--forming activity
is not over yet, although partly hidden by dust.

As for radio galaxies in the cluster, Bacchi et al. (\cite{bac03})
confirm the presence of three radio galaxies at the cluster center of
A2219 (see Owen et al. \cite{owe92}) identified in our catalogue
with no. 51, 77, 85 (i.e. S3, S4, and S6 respectively in Fig.~9 by
Bacchi et al. \cite{bac03}). The northernmost galaxy no. 77, which
lies in the low--velocity tail of the cluster velocity distribution
($\rm{v}=62364$ \kss), shows a tailed structure pointing towards NW
direction, with a sharp bend towards the NE (see Owen \& Ledlow
\cite{owe97}). The galaxy no. 85 is the second brightest galaxy of the
cluster ($\sim$ 17 R mag) and lies at $<1\arcmin$ SE separation from
the cD galaxy.

\section{Discussion and conclusion}

We analyze the internal dynamics of A2219 on the basis of
spectroscopic data for 132 galaxies in a cluster region of $\sim
5\arcmin$ ($\sim 0.8$ \h at the cluster distance) centered on the cD
galaxy. Out of 132 spectra, 27 galaxies come from new observations
carried out at the TNG.  We find that A2219 appears as a peak in the
velocity space at $z=0.225$, and select 113 cluster members.  The
investigation of the dynamical status is also performed by using
X--ray data stored in the Chandra archive. Further valuable
information comes from other bands -- optical photometric, infrared,
and radio data -- which are analyzed and/or discussed, too.

The value we obtain for the line--of--sight velocity dispersion
$\sigma_{\rm v}\sim 1400$ \ks is in the high--tail of the
distribution of cluster velocity dispersions (see Fadda et
al. \cite{fad96}; Mazure et al. \cite{maz96}; Girardi \& Mezzetti
\cite{gir01}).  This global value of $\sigma_{\rm v}$ is consistent
with the average value of $T_{\rm X} \sim 10$ keV coming from the
X--ray analysis ($\beta_{\rm{spec}}$=1.10$^{+0.15}_{-0.12}$, see
also Fig.~\ref{figprof}) and with the value of $L_{\rm X}(0.1-2.4\
\rm{keV})\sim 5\;h^{-2}\;10^{44}$ erg s$^{-1}$ (Ebeling et
al. \cite{ebe96}), converted in $L_{\rm{X,bol}}\sim2\cdot 10^{45} \
h^{-2}$ erg s$^{-1}$ (see $L_{\rm X}$--$\sigma_{\rm v}$ relation by,
e.g., Wu et al. \cite{wu99}, and Girardi \& Mezzetti \cite{gir01}).

Therefore, on the basis of global properties only, we would conclude
that A2219 is not far from the dynamical equilibrium and would trust
the large virial mass estimate $M(\sim 2.2$\,\hh$)\sim
2.8\times10^{15}$ \msun.  However, the high values of $\sigma_{\rm
v}$, $T_{\rm X}$, and $L_{\rm X}$ could be due to the expected
increase during a cluster--merging phase (e.g., Schindler \&
B\"ohringer \cite{sb93}; Schindler \& M\"uller \cite{sm93}). Their
agreement could be a mere coincidence, and our virial mass estimate an
overestimate of the true value.

We find observational evidence that the cluster is not
dynamically relaxed.  Our optical and X--ray analyses show that A2219
is elongated in the SE--NW direction. This elongation is seen in: the
spatial distribution of the colour--selected likely cluster members;
the shape of the cD galaxy; the X--ray contour levels of the Chandra
image; the gradient in the velocity dispersion and in the X--ray
temperature. In particular the ellipticity of X--ray isophotes
is significantly higher than 0 and corresponds to an axis-ratio=0.66,
which is much lower than the usual value for galaxy clusters (median
axis-ratio with 99\% c.l. errors=$0.82^{+0.04}_{-0.02}$ from Mohr et
al. \cite{moh95}).  Note that Dahle et al. (\cite{dah02}) found that
the mass distribution of this cluster recovered from a weak lensing
analysis appears even more elongated than the light and galaxy number
density distribution.

Very interestingly, the optical data show a somewhat smaller PA than
X--ray data ($\sim 110\degr$ and $\sim 130\degr$, respectively).  A
different direction in the elongation of light and hot gas
distribution could be the result of a cluster merger in an advanced
phase as shown by numerical simulations where collisional and
collisionless components react in different way (e.g., Ricker \&
Sarazin \cite{ric01}; Schindler \cite{sch02}). In fact, the
evidence of an ongoing merger in the cluster core was first suggested
by Smail et al. (\cite{sma95}) to explain the difference they find in
the orientation of the mass and gas distributions (PA=115 and 128
degrees), as obtained from gravitational lensing analysis and
ROSAT/HRI X--ray data, respectively.

As for the presence of the velocity gradient roughly in the E--W
direction, this suggests that the structures are moving into the plane
of the sky at $\sim 45\degr$ (e.g., Quintana et al. \cite{qui96};
Roettiger \& Flores \cite{roe00}).  In fact, intermediate angles are
the most suitable to allow detection of the cluster elongation and
contemporary the velocity gradient during a merging phase.

The hypothesis of an advanced--phase merging is also supported by the
lack of a cool center in the cluster (see also Allen \& Fabian
\cite{all98}), as expected since strong cluster mergers disrupt the
cool cores (Peres et al. \cite{per98}; Roettiger et al. \cite{roe96}),
and the existence in the core of several star--forming or
post--starburst galaxies, whose activity could be produced by the
cluster merger (e.g., Bekki \cite{bek99}). In particular, three of these
active galaxies are part of the filament at NW of the cD, also visible
as a cool component in the X--ray softness ratio map (see
Fig.~\ref{figxy} and Fig.~\ref{figsoftness}), suggesting that this
structure might be already involved in a merger event.  Moreover, the
shape of the radio tail of the radio galaxy no. 77 indicates that
the velocity of this galaxy was originally pointing towards the cluster
center (cD galaxy), but has rather suddenly changed the motion
direction, probably because of the cluster merger. 

The structure we find in the cluster core could be related to the
cluster merger, too. Considering optical data, the Dressler--Schectman
test shows the possible existence of a substructure located at SE,
very close to the cD galaxy.
As for X--ray data, our wavelet multi--scale analysis puts in evidence
three significant emission peaks located within 1$\arcmin$ from the
cluster center. Of these, two peaks are possibly connected to
intrinsic X--ray emission from the cD and the active galaxy no. 81,
but no optical counterpart is found for the third, intermediate peak.
Our wavelet X--ray analysis also detects an external diffuse
substructure located 2$\arcmin$ SE of the cD galaxy.  We suggest that
this structure is a premerging clump.  In fact, it coincides with the
cold SE X--ray region and thus could be well explained by the
two--dimensional superposition of a cold clump in the foreground of
the main cluster.  The presence of this foreground clump could also
explain why, on average, the velocity at the East is lower than in the
West since at least for some galaxies the low velocity is of
cosmological rather than of kinematic origin.  These anomalous low
velocity galaxies could also produce a spurious enhancement of the
observed velocity dispersion in the SE region (see Table~2).  A
population of very close foreground galaxies could also explain why
{\em very red} galaxies strangely differ from {\em not--very red}
galaxies both in kinematic and spatial properties (see Sect.~3.3). In
fact, while the former are really associated with the main structure
of A2219, the latter could be mixed with the clump members, i.e. a
population of foreground -- and thus less red -- galaxies.
Unfortunately, we have very poor redshift information in the SE region
to confirm our hypothesis and identify clump members.

On a much larger scale, the cluster is not so well isolated in the
redshift space, and a next possible merging clump might be the one at
$z\sim 0.19$ (see Fig.~\ref{figden}).

Summarizing the above evidence, we suggest a complex merging scenario
for A2219 involving many clumps, possibly in different dynamical
states. We suggest that the main, original structure (hosting the cD
galaxy) is subject to an ongoing merger with few clumps aligned in a
foreground filament obliquely oriented with respect to the
line--of--sight; the filament projected on the sky lies in the SE--NW
direction. In this scenario, the merging with one or more clumps are
already in a very advanced phase, well after the first core passage,
while one or more clumps are still in a pre--merging phase.

In the context of the above scenario, we argue that the presence
of a radio--halo in A2219 is related to the existence of an ongoing,
very advanced phase of merging between the main cluster and one or
more groups.
In fact, the elongation of radio map contours (see
Fig.~\ref{figisofote}) has the same direction of the X--ray surface
brightness contours. Moreover, the time necessary to re--accelerate
the electrons producing the radio--halo as a consequence of the
cluster merger ($\sim$1 Gyr) is comparable to that needed to obtain
post--starburst spectral signatures in galaxies of which A2219 is rich
(see Sect.~5).

In conclusion, our scenario supports the view of the connection
between extended radio emission and merging phenomena in galaxy
clusters.

\begin{acknowledgements}
We thank Luca di Fabrizio for B and R calibration observations at the
TNG telescope and the anonymous referee for useful suggestions and
comments. We particularly thank Ian Smail for providing us accurate
Palomar 5--m telescope photometry.  Work partially supported by the
Italian Ministry of Education, University, and Research (MIUR, grant
COFIN2001028932 "Clusters and groups of galaxies, the interplay of
dark and baryonic matter"), by the Italian Space Agency (ASI), and by
INAF (Istituto Nazionale di Astrofisica) through grant D4/03/IS.  This
publication makes use of data accessed as Guest User at the Canadian
Astronomy Data Center, which is operated by the Dominion Astrophysical
Observatory for the National Research Council of Canada's Herzberg
Institute of Astrophysics (http://cadcwww.dao.nrc.ca/cfht/cfht.html),
as well as of data obtained from the Chandra data archive at the NASA
Chandra X--ray center (http://asc.harvard.edu/cda/).
\end{acknowledgements}

\end{document}